\documentclass[sigconf]{acmart}

\usepackage{booktabs} 
\usepackage{microtype}
\usepackage[latin1]{inputenc}
\usepackage[OT2,T1]{fontenc}
\usepackage[linesnumbered,ruled,vlined]{algorithm2e}
\usepackage{algpseudocode}
\usepackage{bm}
\usepackage{amsmath}
\usepackage{url}
\usepackage{psfrag}
\usepackage{balance}
\usepackage{subfigure}
\usepackage{graphicx}
\usepackage{float}
\usepackage{multirow}
\usepackage{pstricks}
\usepackage{array}
\usepackage{enumitem}
\usepackage{hyperref}
\usepackage{pifont}
\usepackage{textcomp}
\usepackage{xcolor}
\graphicspath{{figs/}}
\def\BibTeX{{\rm B\kern-.05em{\sc i\kern-.025em b}\kern-.08em
    T\kern-.1667em\lower.7ex\hbox{E}\kern-.125emX}}

\newtheorem{theorem}{\textbf{Theorem}}
\newenvironment{pf}{{\it \textbf{Proof.} }}{\hfill $\square$\medskip}

\copyrightyear{2024}
\acmYear{2024}
\setcopyright{acmlicensed}\acmConference[KDD '24]{Proceedings of the 30th ACM SIGKDD Conference on Knowledge Discovery and Data Mining}{August 25--29, 2024}{Barcelona, Spain}
\acmBooktitle{Proceedings of the 30th ACM SIGKDD Conference on Knowledge Discovery and Data Mining (KDD '24), August 25--29, 2024, Barcelona, Spain}
\acmDOI{10.1145/3637528.3671695}
\acmISBN{979-8-4007-0490-1/24/08}


\begin{document}

\title{QSketch: An Efficient Sketch for Weighted Cardinality Estimation in Streams}

\author{Yiyan Qi$^*$}
\affiliation{
   International Digital Economy Academy (IDEA)\\
  \city{Shenzhen}
  \country{China}
}
\email{qiyiyan@idea.edu.cn}

\author{Rundong Li$^*$}
\affiliation{%
    MOE KLINNS Lab\\
   Xi'an Jiaotong University\\
  \city{Xi'an}
  \country{China}
}
\email{xjtulirundong@stu.xjtu.edu.cn}

\author{Pinghui Wang$^\ddag$}
\affiliation{%
    MOE KLINNS Lab\\
  Xi'an Jiaotong University\\
  \city{Xi'an}
  \country{China}
}
\email{phwang@xjtu.edu.cn}

\author{Yufang Sun}
\affiliation{%
MOE KLINNS Lab\\
   Xi'an Jiaotong University\\
  \city{Xi'an}
  \country{China}
}
\email{sunyufang00@stu.xjtu.edu.cn}

\author{Rui Xing}
\affiliation{%
MOE KLINNS Lab\\
  Xi'an Jiaotong University\\
  \city{Xi'an}
  \country{China}
}
\email{xingrui128719@163.com}

\thanks{$^*$ Equal Contribution.}
\thanks{$^\ddag$ Corresponding Author.}
\renewcommand{\shortauthors}{Yiyan Qi et al.}

\begin{abstract}
Estimating cardinality, i.e., the number of distinct elements, of a data stream is a fundamental problem in areas like databases, computer networks, and information retrieval.
This study delves into a broader scenario where each element carries a positive weight. 
Unlike traditional cardinality estimation, limited research exists on weighted cardinality, with current methods requiring substantial memory and computational resources, challenging for devices with limited capabilities and real-time applications like anomaly detection.
To address these issues, we propose QSketch, a memory-efficient sketch method for estimating weighted cardinality in streams. QSketch uses a quantization technique to condense continuous variables into a compact set of integer variables, with each variable requiring only 8 bits, making it 8 times smaller than previous methods. Furthermore, we leverage dynamic properties during QSketch generation to significantly enhance estimation accuracy and achieve a lower time complexity of $O(1)$ for updating estimations upon encountering a new element. Experimental results on synthetic and real-world datasets show that QSketch is approximately 30\% more accurate and two orders of magnitude faster than the state-of-the-art, using only $1/8$ of the memory.
\end{abstract}
\maketitle

\section{Introduction} \label{sec:introduction}
Real-world systems generate data in a streaming fashion.
Examples range from financial transactions to Internet of Things (IoT) data, network traffic, call logs, trajectory logs, etc.
Computing the cardinality, i.e., the number of distinct elements, of such a stream is fundamental in research areas like databases, machine learning, and information retrieval.
For example, online games and mobile apps usually use \textit{daily active users} (DAU), i.e., the number of distinct active users within a day, as a metric to measure the level of engagement.
Other examples include network security monitoring \cite{Estan2003} and connectivity analysis in the Internet graph \cite{palmer2001connectivity}.

Due to the unknown or even unlimited size and the high-speed nature of these data streams,
it is infeasible to collect the entire data when the computation and memory resources of data collection devices (e.g., network routers) are limited.
To solve this challenge, considerable attention has been paid to designing fast and memory-efficient cardinality estimating algorithms via sketching techniques \cite{Whang1990, Durand2003, FlajoletAOFA07, GiroireDAMNew2009}.
They build a compact data summary (i.e., sketch) on the fly and then estimate the cardinality from the generated sketch.
The above cardinality computing problem can be generalized to a weighted scenario, where each element $\mathrm{e}$ in the data stream is associated with a positive weight $w \in \mathbb{R}_+$.
In this new scenario, the goal is to compute the total sum of weights for all distinct elements, i.e., weighted cardinality.
The weighted cardinality has various applications, including 1) In database systems, an example is a SQL query like ``SELECT DISTINCT * FROM TABLE''. 
In addition to the query's cardinality, understanding the total size of the resultant set aids in optimizing performance and managing resources~\cite{long2005three,ozcan2011cost}.
Here, the weighted cardinality represents the total size (in bytes) of the query result, calculated as the cumulative size of all distinct records, weighted by their row size. 2) In a voting system, each voter may have a weight based on their expertise and we need it to figure out the final voting result. 3) In an app or website, users with more activity might be assigned higher weights. This metric allows for a more nuanced understanding of the app other than Daily Active User (DAU). Weighted cardinality computation is important in scenarios where individual contributions vary in importance.

Despite the plenty of works for estimating the regular cardinality, 
little attention has been paid to the problem of \textbf{W}eighted \textbf{C}ardinality \textbf{E}stimation (WCE).
Lemiesz \cite{lemiesz2021algebra} conducts a formal study on the WCE problem.
The proposed method maps each element in the data stream into $m$ exponential variables concerning the element's weight.
To guarantee estimation accuracy, 
$m$ is set to hundreds or thousands, making it infeasible to deal with real-time streams.
Zhang et al. \cite{zhang2023fast} proposed a method \textit{FastGM} to decrease the update time complexity of Lemiesz's method.
Instead of generating $m$ variables independently, 
FastGM generates those exponential variables in ascending order and early stops the generation when the value is greater than the maximal value stored in the current registers.
A recent method FastExp Sketch~\cite{lemiesz2023efficient} shares the same idea with FastGM.
The above three methods use 64-bit floating-point registers to store these exponential variables. 
When a large value of $m$ is employed for improved accuracy, it becomes memory-intensive for devices with limited computational and storage resources.
Additionally, they require $O(m)$ operations to estimate weighted cardinality, making them computationally expensive when aiming to provide anytime-available estimation for real-time applications.

We develop a memory-efficient sketch, \emph{QSketch}, to estimate the weighted cardinality of distinct elements in a data stream.
QSketch generates $m$ independent exponential variables for each incoming element in descending order. 
This process is employed to update the $m$ \textbf{integer} registers, 
and an early termination occurs when a generated variable is smaller than the values in all registers.
QSketch employs a novel mapping strategy that transforms continuous exponential variables into discrete variables, using a small set of integer registers to represent data streams with varying weighted cardinalities. 
Consequently, each register in our QSketch requires no more than $5$ bits, making it up to $13 \times$ smaller than both Lemiesz's method and FastGM.
To reduce time cost and estimation error, we propose an extension of QSketch, \textit{QSketch-Dyn}, to monitor the weighted cardinality on the fly.
QSketch-Dyn shares the same data structure as QSketch but only needs to compute one variable for each element.
It utilizes the dynamic property of our QSketch to reduce estimation error significantly. 
We summarize our main contributions as:

\begin{itemize}[leftmargin=*]
\item We propose a memory-efficient sketch method QSketch to estimate the weighted cardinality of distinct elements in a data stream. 
QSketch employs a novel mapping strategy that transforms continuous exponential variables into discrete variables, using a small set of integer registers to represent data streams with varying weighted cardinalities.
\item We also present QSketch-Dyn, an advanced variant of QSketch,
which leverages the inherent dynamic nature of QSketch, enabling real-time tracking of weighted cardinality. 
\item We conduct experiments on both synthetic and real-world datasets. The experimental results demonstrate that our new sketching method achieves approximately 30\% more accurate and is two orders of magnitude faster than the state-of-the-art while requiring only 1/8 of the memory usage.

\end{itemize}

The rest of this paper is organized as follows.
Section~\ref{sec:problem} introduces the problem.
Section~\ref{sec:preliminaries} briefly discusses preliminaries.
Section~\ref{sec:method} presents our method QSketch and QSketch-Dyn.
The performance evaluation and testing results are presented in Section~\ref{sec:results}.
Section~\ref{sec:related} summarizes related work. Concluding remarks then follow.

\section{Problem Formulation} \label{sec:problem}
We first introduce some notations.
Let $\Pi=\mathrm{e}^{(1)} \cdots \mathrm{e}^{(t)} \cdots$ denote a data stream, 
where an element $\mathrm{e}^{(t)}$ arriving at time $t$ corresponds to one of the elements $x_1, \ldots, x_n$ and each $x_i, 1\le i \le n$ has a positive weight $w_i>0$.
Note that an element $\mathrm{e}$ may appear multiple times in the stream.
Denote $N_{\Pi}^{(t)}$ as the set of distinct elements that occurred in stream $\Pi$ before and including time $t$.
Then, the weighted cardinality of stream $\Pi$ at $t$ is defined as 
\begin{equation}
C^{(t)}=\sum_{x_i \in N_{\Pi}^{(t)}} w_i.
\end{equation}
This paper aims to develop a sketch method to estimate the weighted cardinality $C^{(t)}$ accurately and efficiently.
We omit the superscript $(t)$ when no confusion arises.

\section{Preliminaries}\label{sec:preliminaries}
In this section, we introduce two existing methods to estimate weighted cardinality and discuss their shortcomings.

\subsection{Existing Methods For WCE}

\begin{algorithm}[t]
\SetKwFunction{continue}{continue}
\SetKwInOut{Input}{input}
\SetKwInOut{Output}{output}
\Input{stream $\Pi$, $m$.}
\Output{sketch $R$.}
\BlankLine
$R \gets [+\infty, \ldots, +\infty]$\;
\ForEach {$(x, w) \in \Pi$}{s
    \For {$j=1,\ldots,m$}{
    	$r_j \gets -\frac{\ln h_j(x)}{w}$ \;
        $R[j]\gets \min(R[j], r_j)$\;
    }
}
\caption{Pseudo-code of Lemiesz's method.}\label{alg:Lemiesz}
\end{algorithm}

\noindent $\bullet$ \textbf{Lemiesz's Method}~\cite{lemiesz2021algebra}
builds a sketch consisting of $m$ registers $R[1], \dots, R[m]$.
Typically, $m$ is set to be thousands to guarantee the desired accuracy.
All $m$ registers are initialized to $+\infty$.
For each $j\in \{1, \ldots, m\}$, let $R^{(t)}[j]$ denote the value of $R[j]$ at time $t$. 
For each $\mathrm{e}^{(t)}$ arriving at time $t$, Lemiesz's method maps it into all $m$ registers independently and each register is updated as
$$
R^{(t)}[j] \leftarrow \min(R^{(t-1)}[j], r_j(\mathrm{e}^{(t)})).
$$
Without loss of generality, we let $\mathrm{e}^{(t)} = x_i$ with weight $w_i$, $1 \le i \le n$.
In the above equation, $r_j(\mathrm{e}^{(t)})$ is defined as
$$
r_j(\mathrm{e}^{(t)}) = -\frac{\ln{h_j(x_i)}}{w_i},
$$
where $h_j(x)$ is a hash function that maps $x$ to $(0, 1)$ uniformly, i.e., $h_j(x) \sim \text{Uniform}(0,1)$, $1 \le j \le m$.
The pseudo-code of Lemiesz's method is shown in Algorithm~\ref{alg:Lemiesz}.
We note that $r_j(\mathrm{e}^{(t)})$ follows an exponential distribution $\text{EXP}(w^{(t)})$ and $R^{(t)}[j]=\min\limits_{x_i \in N^{(t)}_{\Pi}} r_j(x_i)$ follows an exponential distribution $\text{EXP}(C^{(t)})$,
in which $C^{(t)}$ is the weighted cardinality at time $t$.
Thus, the summation variable $G_m=\sum^{m}_{j=1} R^{(t)}[j]$ follows a gamma distribution $G_m\sim\Gamma(m, C^{(t)})$.
According to~\cite{witkovsky2001computing}, the inverse of the summation variable $G_m$, i.e., $1/G_m$,
has the inverse gamma distribution $1/G_m\sim\Gamma^{-1}(m, C^{(t)})$,
and we have
\[
\mathbb{E}[1/G_m]=\frac{C^{(t)}}{m-1},\quad
\mathbb{V}\text{ar}[1/G_m] = \frac{(C^{(t)})^2}{(m-2)(m-1)^2},
\]
where the first equation holds for $m\ge 2$ and second for $m\ge 3$.
Then at time $t$, Lemiesz gives an unbiased estimator of $C^{(t)}$ as 
\begin{equation}\label{eq:est1}
    \hat{C}^{(t)} = \frac{m-1}{\sum_{j=1}^m R^{(t)}[j]}.
\end{equation}
The above estimator's variance is computed as
$$
\mathbb{V}\text{ar}[\hat{C}^{(t)} / C^{(t)}] = \frac{1}{m-2}.
$$

\noindent $\bullet$ \textbf{FastGM.}
Lemiesz's method needs $O(m)$ time to update an element
In practice, $m$ is usually thousands to achieve the expected accuracy and is infeasible for high-speed streams.
To solve this problem, Zhang et al.~\cite{zhang2023fast} proposed FastGM,
reducing the time complexity from $O(m)$ to $O(1)$.
FastGM shares the same sketch structure as Lemiesz's method.
The difference is that $m$ random variables in FastGM $-\frac{\ln{h_1(x_i)}}{w_i}$, $\ldots$, $-\frac{\ln{h_m(x_i)}}{w_i}$ are generated in ascending order as $\left( -\frac{\ln{h_{\pi_1}(x_i)}}{w_i}, \pi_1\right)$, $\ldots$, $\left( -\frac{\ln{h_{\pi_m}(x_i)}}{w_i}, \pi_m\right)$, where $-\frac{\ln{h_{\pi_1}(x_i)}}{w_i} < \cdots < -\frac{\ln{h_{\pi_m}(x_i)}}{w_i}$ and $\pi_1, \ldots, \pi_m$ is a random permutation of integers $1, \ldots, m$.
Once the current obtained random variable $-\frac{\ln{h_{j_1}(x_i)}}{w_i}$ is larger than all values in registers $R^{(t)}[1], \ldots, R^{(t)}[m]$, 
there is no need to generate the following random variables because they have no chance to change the sketch.
In detail, for each $\mathrm{e}^{(t)}=x_i$, $1 \le i \le n$, FastGM generates the first exponential variable as
\begin{equation}\label{equ:expv1}
    r_{\pi_1}(\mathrm{e}^{(t)}) = - \frac{1}{m} \cdot\frac{\ln h_{\pi_1}(x_i)}{w_i},
\end{equation}
and the following exponential variables $r_{\pi_j}(\mathrm{e}^{(t)}), 2 \le j \le m$ are generated in ascending order as 
\begin{equation}\label{equ:expv2}
    r_{\pi_j}(\mathrm{e}^{(t)}) = r_{\pi_{j-1}}(\mathrm{e}^{(t)}) - \frac{1}{m - j + 1} \cdot \frac{\ln h_{\pi_j}(x_i)}{w_i}.
\end{equation}
After generating a hash value $r_{\pi_j}(\mathrm{e}^{(t)})$, $j\in \{1, \ldots, m\}$, 
FastGM uses the Fisher-Yates shuffle~\cite{fisher1938statistical} to find its position $\pi_j\in\{1,\ldots,m\}$.
Specially, let $(\pi_1, \ldots, \pi_m)$ be initialized to $(1, \ldots, m)$.
To obtain the position of the $j$-th smallest hash value, 
FastGM randomly chooses a position $k \in \{j,j+1, \ldots, m\}$, 
swaps $\pi_k$ and $\pi_j$, and updates $R^{(t)}[\pi_j]$ with $r_{\pi_j}(\mathrm{e}^{(t)})$.
FastGM uses an extra register $r^*$ to perform an early stop to record the maximal values among $m$ registers.
Register values in FastGM follow the same distribution as LM.
They share the same weighted cardinality estimator and the same estimation errors.
Besides, we find that a recent method FastExpSketch~\cite{lemiesz2023efficient} shares the same idea with FastGM.

\subsection{Limitations of Existing Methods}

\noindent $\bullet$ \textbf{Memory-consuming.}
Lemiesz's method and FastGM use 32-bit or 64-bit floating-point registers to store hash variables.
They require $32m$ or $64m$ bits of memory to store $m$ hash variables for a stream $\Pi$.
When we need a large $m$ for better accuracy or there are many different streams, 
it is memory-intensive for devices (e.g. IoT devices or routers) with limited computational and storage resources.
Therefore, reducing the number of bits for each register is practical, 
saving storage space and improving the computational efficiency for sketch operations~\cite{lemiesz2021algebra, li2010b}.

\noindent $\bullet$ \textbf{Time-consuming.}
Lemiesz's method requires updating each register for element insertion, incurring an $O(m)$ time cost. 
While FastGM improves upon this by ordering register updates, 
it still demands $O(m)$ time in the worst-case scenario when element weights grow over time. 
Furthermore, the time complexity for the estimation process in both methods is quantified as $O(m)$.
Such a computational demand poses challenges for providing consistent, real-time estimations in applications that require immediate data availability.

\section{Our Method}\label{sec:method}
In this section, we first introduce a compact sketch QSketch (\textbf{Q}uan-tization \textbf{Sketch}) which utilizes the quantization technique (i.e., mapping continuous infinite values to a small set of discrete values).
We design a novel estimator to estimate weighted cardinality.
Then, we exploit the dynamic properties of the register arrays over time to significantly improve the estimation accuracy and reduce the time cost to monitor the weighted cardinality on the fly. 

\subsection{Basic Idea}
\begin{figure}
    \centering
    \includegraphics[width=0.49\textwidth]{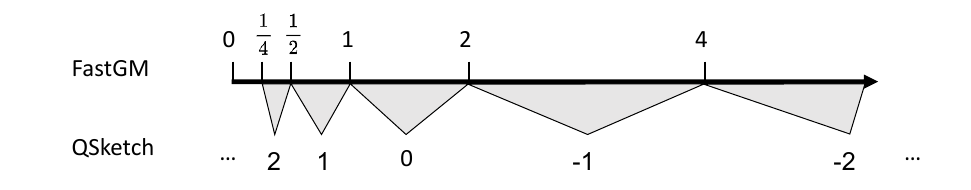}
    \vspace{-5mm}
    \caption{Basic idea of QSketch}
    \vspace{-5mm}
    \label{fig:basic_idea}
\end{figure}
QSketch reduces the size of 32-bit or 64-bit floating-point registers to a smaller bit size (5 or 6 bits) through quantization. 
This process transforms an infinite range of continuous values into a limited set of discrete values. 
Figure~\ref{fig:basic_idea} illustrates how QSketch's hash value mappings compare to those in FastGM and Lemiesz's method.

\subsection{QSketch}\label{sec:qsketch}
\noindent \textbf{Data structure and update procedure.}
Denote $R$ as the sketch with $m$ registers $R[1],\ldots,R[m]$,
and $h_1, \ldots, h_m$ as $m$ independent hash functions, 
each of them mapping $x$ to a random value in range $(0,1)$ uniformly, i.e., $h_j(x) \sim \text{Uniform}(0,1)$, $1 \le j \le k$.
When inserting $\mathrm{e}$ which is associated with i $x$ and weight $w$, we generate $m$ variables $y_1(\mathrm{e}), \ldots, y_m(\mathrm{e})$ as
\begin{equation}\label{eq:quantization}
    y_{j}(\mathrm{e}) = \lfloor -\log_2 ( r_j(\mathrm{e}) ) \rfloor,
\end{equation}
where $r_j(\mathrm{e}) = -\frac{\ln h_{j}(x)}{w}$ is an exponential random variable and $\lfloor \cdot \rfloor$ is the round-down operation.
Then QSketch updates as
\begin{equation}\label{eq:update}
		R[j] \gets \max(R[j], y_j(\mathrm{e})).
\end{equation} 
Following~\cite{zhang2023fast}, 
we generate $m$ variables $r_{\pi_1}(\mathrm{e}), \ldots, r_{\pi_m}(\mathrm{e})$ in an ascending order, 
in which $(\pi_1, \ldots, \pi_m)$ is a random permutation of $(1, \ldots, m)$.
As a result, $y_{\pi_1}(\mathrm{e}),$ $\ldots, y_{\pi_m}(\mathrm{e})$ are generated in a descending order.
The update procedure is shown in Algorithm~\ref{alg:QSketch0}.
We use the Fisher-Yates shuffle \cite{fisher1938statistical} (Line 11-12 in Algorithm~\ref{alg:QSketch0}) to quickly find the position to be updated and use $j^*$ to record the register's index that records the sketch's minimum values.
A brief introduction to the Fisher-Yates shuffle is in Appendix~\ref{alg:fys}.
When the generated variable is smaller than $R[j^*]$, we early stop the generation procedure.

\begin{algorithm}[t]
\caption{Update procedure of QSketch.}\label{alg:QSketch0}
	\SetKwFunction{continue}{continue}
	\SetKwFunction{Swap}{Swap}
	\SetKwFunction{RandInt}{RandInt}
        \SetKwFunction{ReBase}{ReBase}
	\SetKwInOut{Input}{input}
	\SetKwInOut{Output}{output}
	\Input{stream $\Pi$, sketch size $m$, register size $b$.}
	\Output{sketch $R$.}
	\BlankLine
	$r_\text{min} \gets -2^{b - 1}+1$; $r_\text{max} \gets 2^{b - 1} - 1$\; 
        $U \gets [r_\text{min}, \ldots, r_\text{min}]$; $j^{*} \gets 1$\; 
	\ForEach {$\mathrm{e} \in \Pi$}{
        \tcc{element $x$ with weight $w$.}
		$[\pi_1, \ldots, \pi_m] \gets [1, \ldots, m]$ \;
		$r \gets 0$\;  
    		\For {$j=1,\ldots,m$}{
                $r \gets r - \frac{\ln h_j(x)}{w(m - j + 1)}$ \;
    			$y \gets \lfloor - \log_2 (r) \rfloor$ \;
    			\If{$y \le R[j^*]$}{
                    \tcc{early stop}
                        break \;    
                    }
    			$k \gets$ \RandInt$(j, m)$ \;
    			\Swap$(\pi_k, \pi_j)$\;
                \If{$y > R[\pi_l]$}{
                    $R[\pi_l] = \min(\max(y, r_\text{min}), r_\text{max})$\;
                    \If{$\pi_l = j^*$}{
                        $j^* \gets \arg\min_{j=1,\ldots, m} R[j]$ \;
                    }
                        
                }
    	}
	}
\end{algorithm}
\noindent \textbf{Weighted Cardinality Estimation.}
Direct use of the estimator in Equation~(\ref{eq:est1}) yields a large estimation error since the quantization leads to a loss of precision.
Instead, 
we design a new estimator with the help of Maximum Likelihood Estimation (MLE).
Before that, we first derive the probability distribution of a single register $R[j], 1 \le j \le m$.
It has been proved that a register value $R$ (without quantization) follows an exponential distribution $\text{EXP}(C_{\Pi})$~\cite{lemiesz2021algebra, zhang2023fast},
in which $C_{\Pi}$ is the weighted cardinality of stream $\Pi$.
In QSketch, according to Equation~(\ref{eq:quantization}), 
we quantize the continuous register value to discrete values.
Particularly, continuous register values in the range $(2^{-(r+1)}, 2^{-r}]$ will be compressed into a discrete value $r$.
Then the probability distribution of a single register value $R[i]$ is
\begin{equation}\label{eq:pri}
\begin{split}
        Pr(R[j] = r | C_\Pi) &= \int_{2^{-(r+1)}}^{2^{-r}} C_{\Pi} e^{-C_{\Pi} x} dx\\
                     &= e^{-C_{\Pi}\cdot 2^{-(r+1)}} - e^{-C_{\Pi}\cdot 2^{-r}}.
\end{split}
\end{equation}

Given the specific register values $R[1], \ldots, R[m]$, 
we compute the likelihood function for $C_{\Pi}$ is
\begin{equation}\label{eq:likelihood}
\begin{split}
L(C_{\Pi}) = \prod_{j=1}^{m} Pr(R[j] | C_{\Pi}) = \prod_{j=1}^{m} \left(e^{-C_{\Pi}\cdot 2^{-(R[j]+1)}} - e^{-C_{\Pi}\cdot 2^{-R[j]}} \right),
\end{split}
\end{equation}
and the derivative of the likelihood function's logarithm is:
\begin{equation}
	\frac{d (\ln L(C_{\Pi}))}{d C_{\Pi}}  = \sum_{j=1}^{m} 2^{-(R[j]+1)} \cdot \frac{2 - e^{C_{\Pi} \cdot 2^{-(R[j]+1)}}}{e^{C_{\Pi} \cdot 2^{-(R[j]+1)}} - 1}.
\end{equation}
To compute the MLE of $C_{\Pi}$, 
let the above formula equal $0$.
Unfortunately, this equation is too complicated to be solved directly.
So we use the Newton-Raphson method to obtain $\hat{C}_{\Pi}$.
Specially, let
\begin{equation}
f(C_{\Pi}) = \sum_{j=1}^{m} 2^{-(R[j]+1)} \cdot \frac{2 - e^{C_{\Pi} \cdot 2^{-(R[j]+1)}}}{e^{C_{\Pi} \cdot 2^{-(R[j]+1)} - 1}},
\end{equation}
The Newton-Raphson method~\cite{akram2015newton} starts from an initial estimation $\hat{C}_\Pi^{(0)}$ and then repeats the following steps:
\begin{equation}
	\hat{C}_\Pi^{(l+1)} \gets \hat{C}_\Pi^{(l)} - \frac{f(\hat{C}_\Pi^{(l)})}{f'(\hat{C}_\Pi^{(l)})},
\end{equation}
until $\hat{C}_\Pi$ converges, where $f'(\hat{C}_\Pi^{(l)})$ is the derivative of $f(C_\Pi)$ at point $C_\Pi = \hat{C}_\Pi^{(l)}$.
To start the iteration, we initialize $\hat{C}^{(0)}_\Pi$ as
$$
\hat{C}^{(0)}_\Pi = \frac{m-1}{\sum_{1\le j \le m} 2^{-R[j]}}.
$$
Since the register values in QSketch are the quantization of register values in FastGM,
$\hat{C}^{(0)}_\Pi$ can be viewed as an approximation of $\hat{C}_\Pi$,
which is reasonable as an initial value to guarantee convergence.

The above MLE-based estimator provides an asymptotically unbiased estimation.
The approximate variance of the above estimator is based on the Cram\'{e}r-Rao bound~\cite{cramer1999mathematical}.
Specifically, we have $\text{Var}[\hat{C}] \approx \frac{1}{I_{\Pi}(\hat{C})}$, 
where $I_{\Pi}(\hat{C})$ is the observed fisher information given stream data $\Pi$, 
i.e., the negative of the second derivative of the log-likelihood function at $\hat{C}$.
Formally, we have 
\[
\text{Var}[\hat{C}] \approx -\frac{1}{f'(\hat{C}_\Pi^{(l)})}.
\]

Through the quantization, all possible values of $y(\mathrm{e})$ are integers,
i.e., $y(\mathrm{e}) \in \mathbb{Z}$.
In practice, we notice that most values of $y(\mathrm{e})$ are concentrated in a small range and we can truncate these generated variables by $y'(\mathrm{e}) = \min(\max(y(\mathrm{e}), r_\text{min}), r_\text{max})$.
As a result, adjusting the probability distribution of the truncated value in each register is necessary, as shown in Equation (\ref{eq:pri}).
\[
Pr(R[j] | C_{\Pi})=
\left\{
\begin{aligned}
&e^{-C_{\Pi}\cdot 2^{-(r_\text{min}+1)}},\quad R[j] \le r_\text{min}; \\
&1 - e^{-C_{\Pi}\cdot 2^{-r_\text{max}}},\quad R[j] \ge r_\text{max}; \\
&e^{-C_{\Pi}\cdot 2^{-(R[j]+1)}} - e^{-C_{\Pi}\cdot 2^{-R[j]}},\ \text{otherwise.}
\end{aligned}
\right.
\]
By substituting the above probability to Equation (\ref{eq:likelihood}),
we get the weighted cardinality estimator under truncated values.
Note that the estimator only fails to give an unbiased estimation when all register values equal $r_\text{min}$ or $r_\text{max}$
as the likelihood function $L(C_{\Pi})$ becomes monotonous without an extremum. 
Fortunately, in the following theorem, we show that by properly setting $r_\text{min}$ and $r_\text{max}$,  
the failure probability is extremely low.

\begin{theorem}\label{theorem1}
    Let $0 < \varepsilon \ll 1$ be a small positive value. 
    Given a sketch of $m$ registers with minimal value $r_\text{min}$ and maximal value $r_\text{max}$, 
    when $- 2^{(r_\text{min}+1)}\cdot\ln{\epsilon} < C_{\Pi} < -2^{r_\text{max}} \ln(1-\epsilon)$, 
    the register values are not in the discrete set $\{r_\text{min},..., r_\text{max}\}$ within a maximum probability of $2\epsilon$.
\end{theorem}

The proof is in Appendix~\ref{appendix:th1}.
When $r_\text{min}=-127$, $r_\text{max}=127$, and $\varepsilon=0.001$, each register value at time $t$ is not tuncatated with a probability of more than $0.998$ when $8.1 \times 10^{-38} \le C_{\Pi} \le 3.4 \times 10^{35}$.
In this case, we need just $8$ bits to store the register values.

\subsection{QSketch-Dyn}\label{sec:qdyn}
QSketch has the same expected time complexity of $O(m\cdot \ln m + n)$ as FastGM~\cite{qi2020fast,zhang2023fast}.
However, 
it still suffers from the worst time complexity $O(m\cdot n)$ under scenarios in which the weights of the elements increase as time progresses.
In addition, we have to solve the MLE problem whenever figuring out the latest weighted cardinality,  
which costs for real-time estimation.
To improve the estimation efficiency, 
we utilize the dynamic property of the sketch and propose QSketch-Dyn to keep track of the weighted cardinality on the fly.

\noindent \textbf{Data structure and update procedure.}
QSketch-Dyn shares the same data structure,
i.e., a bit array of size $m$,
and the same set of hash functions $h_1,\ldots, h_m$ as Lemiesz's method~\cite{lemiesz2021algebra}.
The main differences are:
1) QSketch-Dyn introduces another hash function $g(x)$, 
which uniformly maps $x$ to some integer in set $\{1,\ldots, m\}$ at random.
2) QSketch-Dyn maintains a tabular $T$ to record the frequency of values in the sketch.
Specifically, the tabular $T$ consists of $2^b$ counters, in which $b$ is the number of bits used by a register.

Next, we introduce how to update an element $\mathrm{e}^{(t)}$ corresponding to $x$ and $w$.
Instead of updating multiple register values as in other methods,
QSketch-Dyn first randomly chooses one register $R[j]$ with $j = g(x)$, $1 \le j \le m$ from the sketch, 
computes its quantized hash value $y_j(\mathrm{e})=\lfloor -\log_2 ( r_j(\mathrm{e}) ) \rfloor$ (Equation~\ref{eq:quantization}),
and update the register value $R[j] \gets \max(R[j], y_j(\mathrm{e}))$ (Equation~\ref{eq:update}).
Once the register value $R[j]$ changes, we then update the tabular $T$ as 
$T[R[j]]\gets T[R[j]]-1$ and $T[y_j(\mathrm{e})]\gets T[y_j(\mathrm{e})]+1$.

\begin{algorithm}[h]
	\SetKwFunction{continue}{continue}
	\SetKwFunction{Swap}{Swap}
	\SetKwFunction{RandInt}{RandInt}
	\SetKwInOut{Input}{input}
	\SetKwInOut{Output}{output}
	\Input{stream $\Pi$, sketch size $m$, register size $b$.}
	\Output{Estimated weighted cardinality $\hat{C}$.}
	\BlankLine
        $r\gets 0$; $q_R\gets 0$; $\hat{C}\gets 0$\;
	$R \gets [r_\text{min}, \ldots, r_\text{min}]$; $T \gets [0, 0, \ldots, 0]$ \;
	\ForEach {$\mathrm{e} \in \Pi$}{
        \tcc{corresponding to element $x$ with weight $w$.}
        $j \gets$  \RandInt$(1, m)$\;
        $r \gets - \frac{\ln h_j(x)}{w}$ \;
        $y \gets \lfloor - \log_2 (r) \rfloor$ \;
        \If{$y > R[j]$}{            
            \If{$T[R[j]-r_\text{min}]>0$}{
                $T[R[j]-r_\text{min}]\gets T[R[j]-r_\text{min}]-1$ \;
                $T[y-r_\text{min}]\gets T[y-r_\text{min}]+1$\;
            }
            \Else{
                $T[y-r_\text{min}]\gets T[y-r_\text{min}]+1$\;
            }
            $R[j]\gets \min(y, r_\text{max})$\;
            $q \gets 0$\;
            \For{ $k=0,\ldots, 2^b-1$}{
                $q\gets q+ T[k] \cdot e^{-w \cdot 2 ^{-(k + r_\text{min} +1)}}$;\
            }
            $q_R = 1 - \frac{q}{m}$\;

            $\hat{C}\gets \hat{C} + \frac{w}{q_R}$\;
        }
        }	
 \caption{Update procedure of QSketch-Dyn.}
 \label{alg:QSketchDyn}
\end{algorithm}

\noindent \textbf{Weighted Cardinality Estimation.}
Denote $q_R^{(t)}$ as the probability of $\mathrm{e}^{(t)}$ changing a register among $R[1], \ldots, R[m]$ at time $t$,
\begin{equation*}
\begin{split}
    q_R^{(t)} &\triangleq \sum\limits_{1 \le j \le m} Pr(y > R[j] \wedge g(x) = j\ |\ R[j])\\
              &= \sum\limits_{1 \le j \le m} Pr(y \ge R[j]+1\ |\ R[j]) \cdot Pr(g(x) = j) \\
              &= \frac{1}{m}\cdot\sum\limits_{1 \le j \le m} \int_{0}^{2^{-(R[j]+1)}} w\cdot e^{-w x} dx \\
              &= 1 - \frac{1}{m} \sum\limits_{1 \le j \le m}  e^{-w\cdot(2^{-(R[j]+1)})}.
\end{split}
\end{equation*}
Let $\hat{C}^{(t)}_\Pi$ denote the weighted cardinality estimate of stream $\Pi$ at time $t$. 
When element $\mathrm{e}$ arrives at time $t$, we update the weighted cardinality estimate as
\begin{equation}\label{eq:est_card}
\hat{C}^{(t)}_{\Pi} \gets \hat{C}^{(t-1)}_{\Pi} + \frac{\textbf{1}(R[j]^{(t)} \neq R[j]^{(t-1)})}{q_R^{(t)}}\cdot w,
\end{equation}
in which $\textbf{1}(R[j]^{(t)} \neq R[j]^{(t-1)})$ is an indicator variable which equals to 1 if element $\mathrm{e}$ changes the register value $R[j]$ (remind that $j = g(x)$) and equals to 0 otherwise.
In later analysis, we will prove that $\hat{C}^{(t)}_{\Pi}$ is an unbiased estimation of ${C}^{(t)}_{\Pi}$.

The computation of probability $q_R^{(t)}$ needs summation over all $m$ registers, 
which is time-consuming when $m$ is set to a large value.
To save time, 
we additionally maintain a tabular $T$ recording the histogram of register values as mentioned above,
where $T[R[j]]$ tracks the count of value $R[j]$ in the current sketch.
As discussed previously, each register occupies $b$ bits and the number of different values is at most $2^b$.
Then, $q_R^{(t)}$ is expressed as
\[
q_R^{(t)} = 1 - \frac{1}{m} \sum\limits_{1 \le j \le 2^b} T[R[j]]\cdot e^{-w\cdot(2^{-(R[j]+1)})}.
\]
We summarize the pseudo-code of the update and estimation procedure of QSketch-Dyn in Algorithm \ref{alg:QSketchDyn}.

\noindent \textbf{Complexity Analysis.}
The time complexity of updating an element for QSketch-Dyn is $O(1)$ since it only chooses one register to update its value.
Then it costs $O(2^b)$ time for QSketch-Dyn to compute $q_R^{(t)}$.
Considering that $b$ is small, this part costs little time.
Finally, QSketch-Dyn tracks the estimated cardinality over time, 
and it costs no time for estimation compared with QSketch.
For space complexity, QSketch uses $m$ registers, 
in which each register occupies $b$ bits.
Besides, QSketch-Dyn maintains a tabular $T$ with $2^b$ counters.
Since there are $m$ registers in the sketch, 
each counter of tabular $T$ occupies at most $\log_2(m)$ bits.
Therefore, the total space complexity is $m\cdot b + 2^b \cdot \log_2(m)$.

\noindent \textbf{Error Analysis.}
For the estimated weighted cardinality $\hat{C}^{(t)}_{\Pi}$ from QSketch-Dyn,
we first prove that our estimator is unbiased and then derive the variance of $\hat{C}^{(t)}_{\Pi}$.
\begin{theorem}\label{theorem4}
     The expectation and variance of $\hat{C}^{(t)}_{\Pi}$ are
     \[
     \mathbb{E}[\hat{C}^{(t)}_{\Pi}] = C^{(t)}_{\Pi},
     \]
     \[
     \text{Var}[\hat{C}^{(t)}_{\Pi}] = \sum\limits_{i\in T_s^{(t)}} (w^{(i)})^2 \mathbb{E}[\frac{1 - q_R^{(i)}}{q_R^{(i)}} ],
     \]
     where $T_s^{(t)}$ is the set of timestamps that each element appears in the stream for the first time.
\end{theorem}

\begin{pf}
See Appendix \ref{appendix:th4}
\end{pf}

\section{Evaluation} \label{sec:results}

All algorithms are implemented in C++ 
and run on a processor with a Quad-Core Intel(R) Xeon(R) CPU E3-1226 v3 CPU 3.30GHz processor.
Our source code is available~\cite{code}.

\subsection{Datasets}
We conduct experiments on both synthetic and real-world datasets.

\noindent $\bullet$ \textbf{Synthetic Datasets.}
We generate synthetic datasets with the following distributions:
Uniform distribution $U(0, 1)$, Gauss distribution $N(1, 0.1)$, and Gamma distribution $\gamma (1, 2)$
For each type of distribution, 
we generate datasets with different sizes of elements, respectively.
The name of the dataset is represented as ``distribution-\#elements''.
For example, Uniform-1k represents the dataset with $1,000$ elements, 
and the weight of each element follows the Uniform distribution $N (0,1)$.
Each dataset is considered a single stream.

\noindent $\bullet$ \textbf{Real-world Datasets.}
We use the following real-world datasets: 
\textbf{Twitter}~\cite{twitter2010},
\textbf{Real-sim}~\cite{Wei2017Consistent}, 
\textbf{Rcv1}~\cite{Lewis2004RCV1}, 
\textbf{Webspam}~\cite{Wang2012Evolutionary},
\textbf{News20}~\cite{keerthi2005modified}
\textbf{Libimseti}~\cite{kunegis2012online}.
Twitter \cite{twitter2010} is a dataset of ``following'' relationships between Twitter users.
The above two datasets are treated as single-stream datasets.
Real-sim \cite{Wei2017Consistent}, Rcv1 \cite{Lewis2004RCV1}, Webspam \cite{Wang2012Evolutionary} 
and News20~\cite{keerthi2005modified} are datasets of web documents from different resources, where each vector represents a document and each entry in the vector refers to the TF-IDF score of a specific word for the document.
Libimseti~\cite{kunegis2012online} is a dataset of ratings of different users, 
where each vector refers to a user and each entry records the user's rating.
These six datasets are considered multi-stream datasets,
and each vector within the dataset is a single stream of elements with a weight.
The details of these datasets are described in Table~\ref{tab:datasets}.
\begin{table}[t]		
	\centering
	\caption{Statistics of real-world datasets.}
        \vspace{-3mm}
		\begin{tabular}{|c|c|c|} \hline
			\textbf{Dataset} &\textbf{\#Documents}&\textbf{Vector Size}\\ \hline
			Real-sim~\cite{Wei2017Consistent}&72,309&20,958\\ \hline
			Rcv1~\cite{Lewis2004RCV1}&20,242&47,236\\ \hline
			Webspam~\cite{Wang2012Evolutionary}&350,000&16,609,143\\ \hline
			News20~\cite{keerthi2005modified}&19,996&1,355,191\\ \hline
			Libimseti~\cite{kunegis2012online}&220,970&220,970\\ \hline
		\end{tabular}
        \vspace{-3mm}
        \label{tab:datasets}
\end{table}

\begin{figure*}[t]
    \centering
    \subfigure[Twitter]{\includegraphics[width=0.329\textwidth, trim=40 5 60 35,clip]{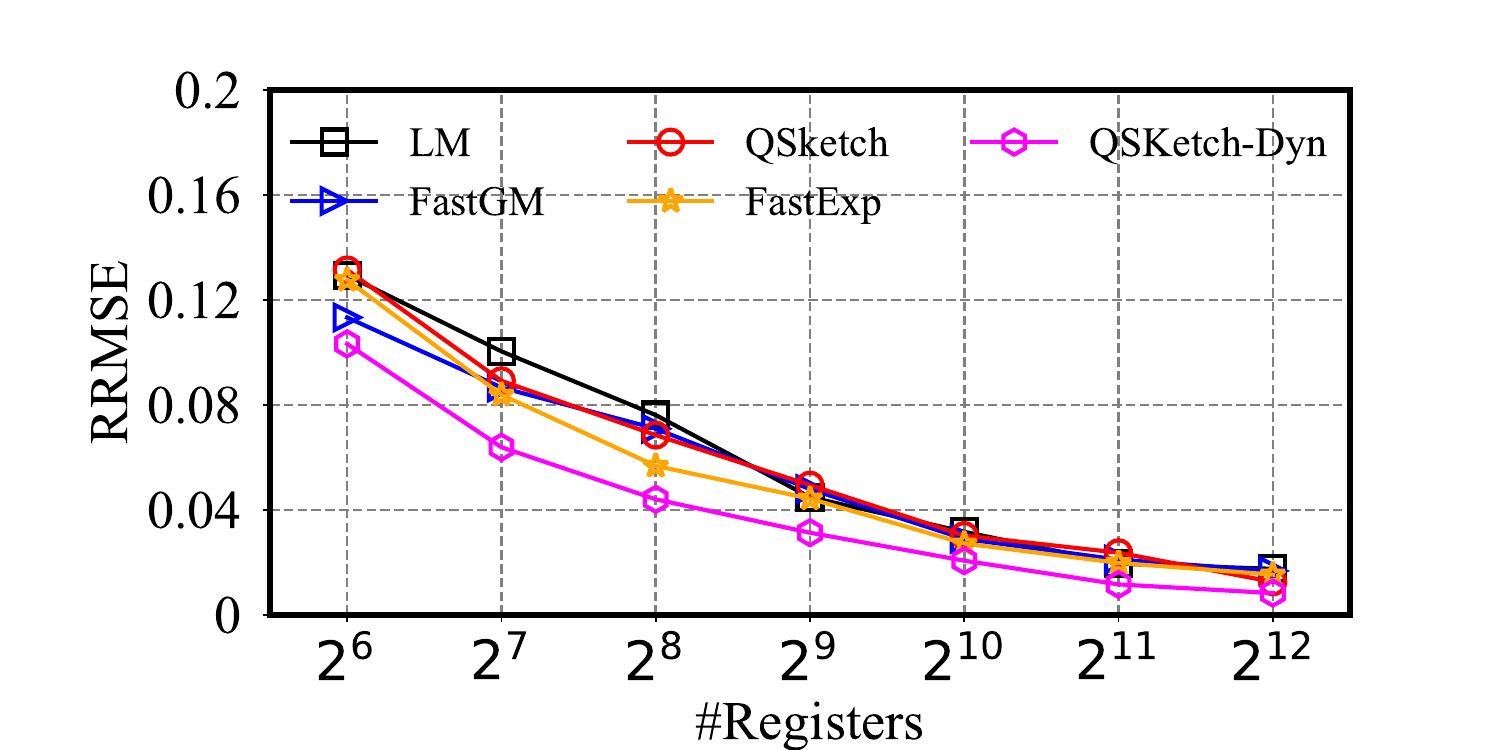}}
    \subfigure[Libimseti]{\includegraphics[width=0.329\textwidth, trim=40 5 60 35,clip]{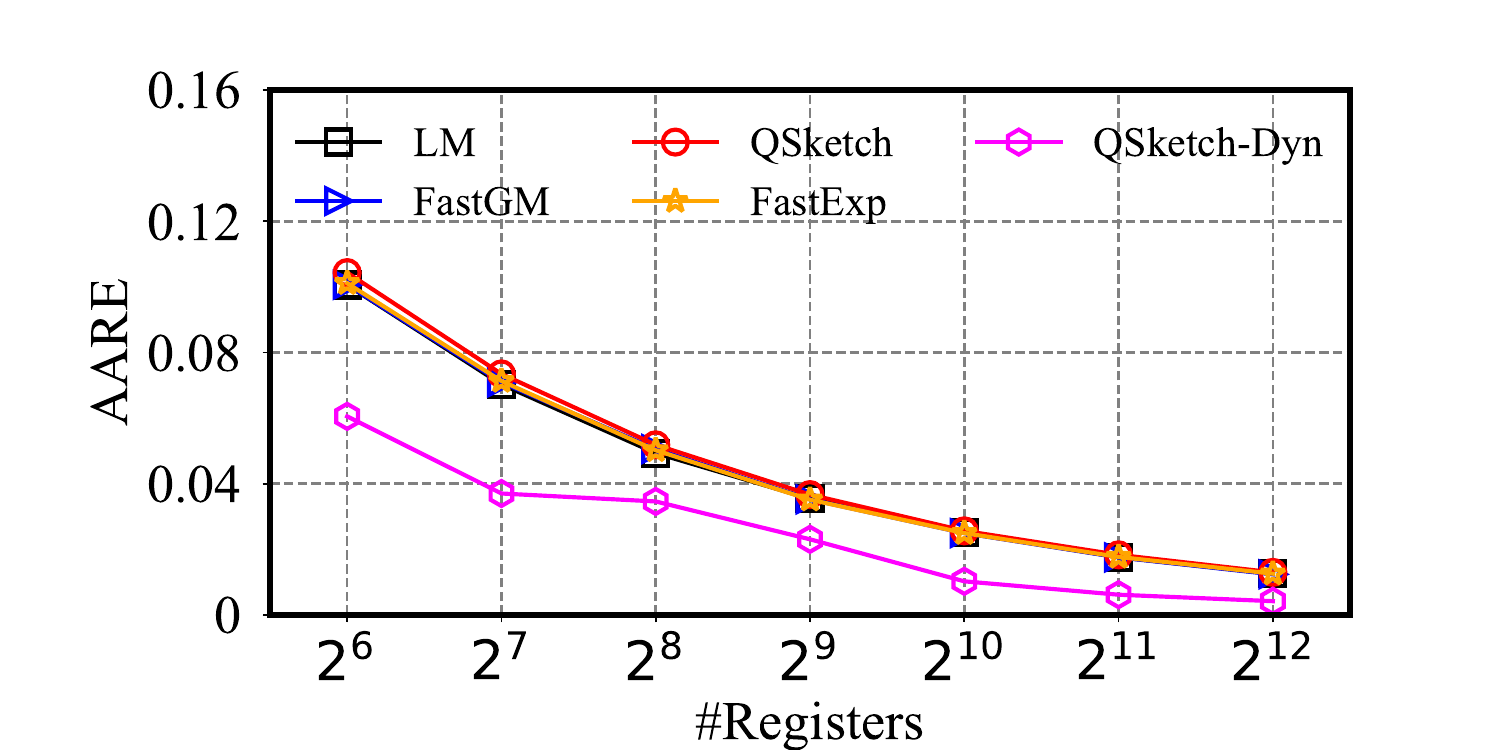}}
    \subfigure[News20]{\includegraphics[width=0.329\textwidth, trim=40 5 60 35,clip]{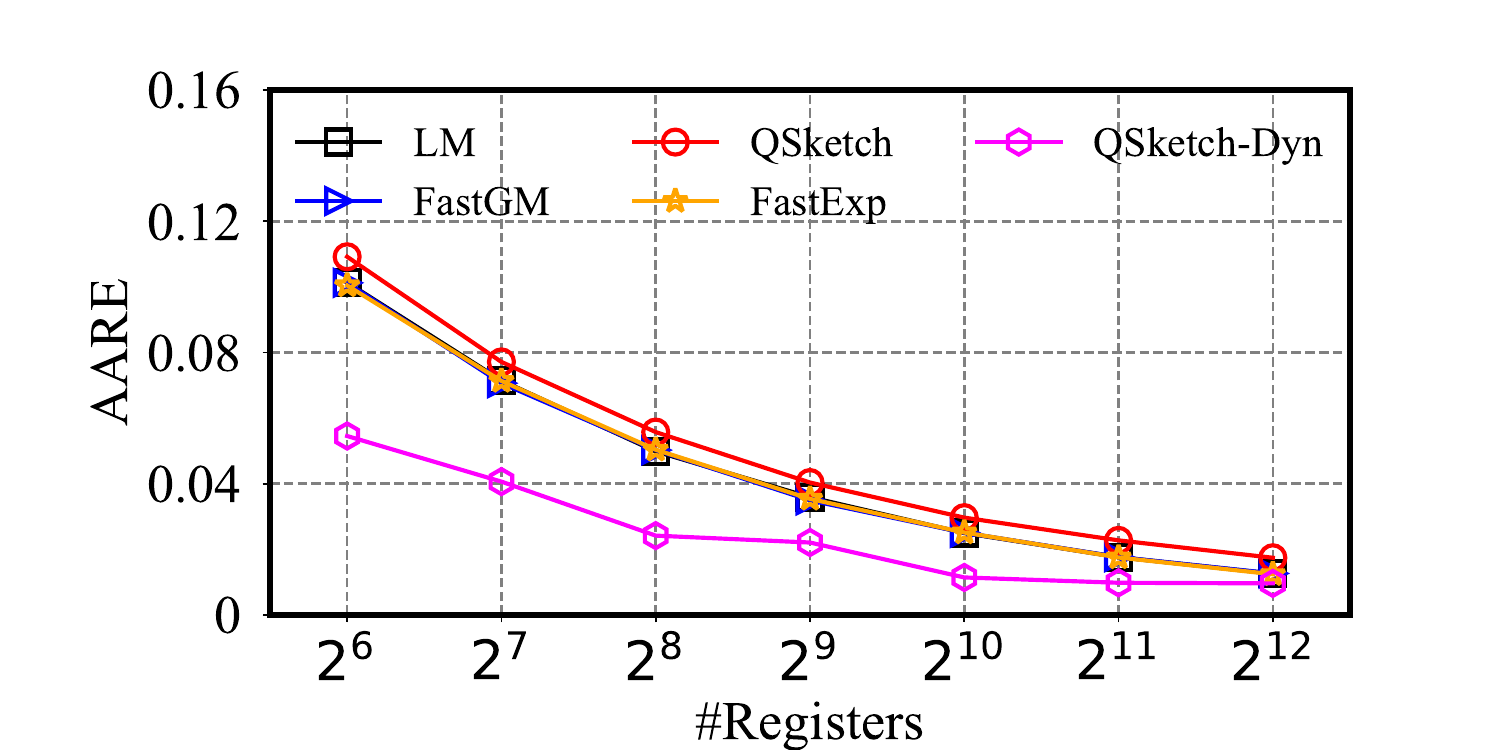}}
    \subfigure[Rcv1]{\includegraphics[width=0.329\textwidth,trim=40 5 60 35,clip]{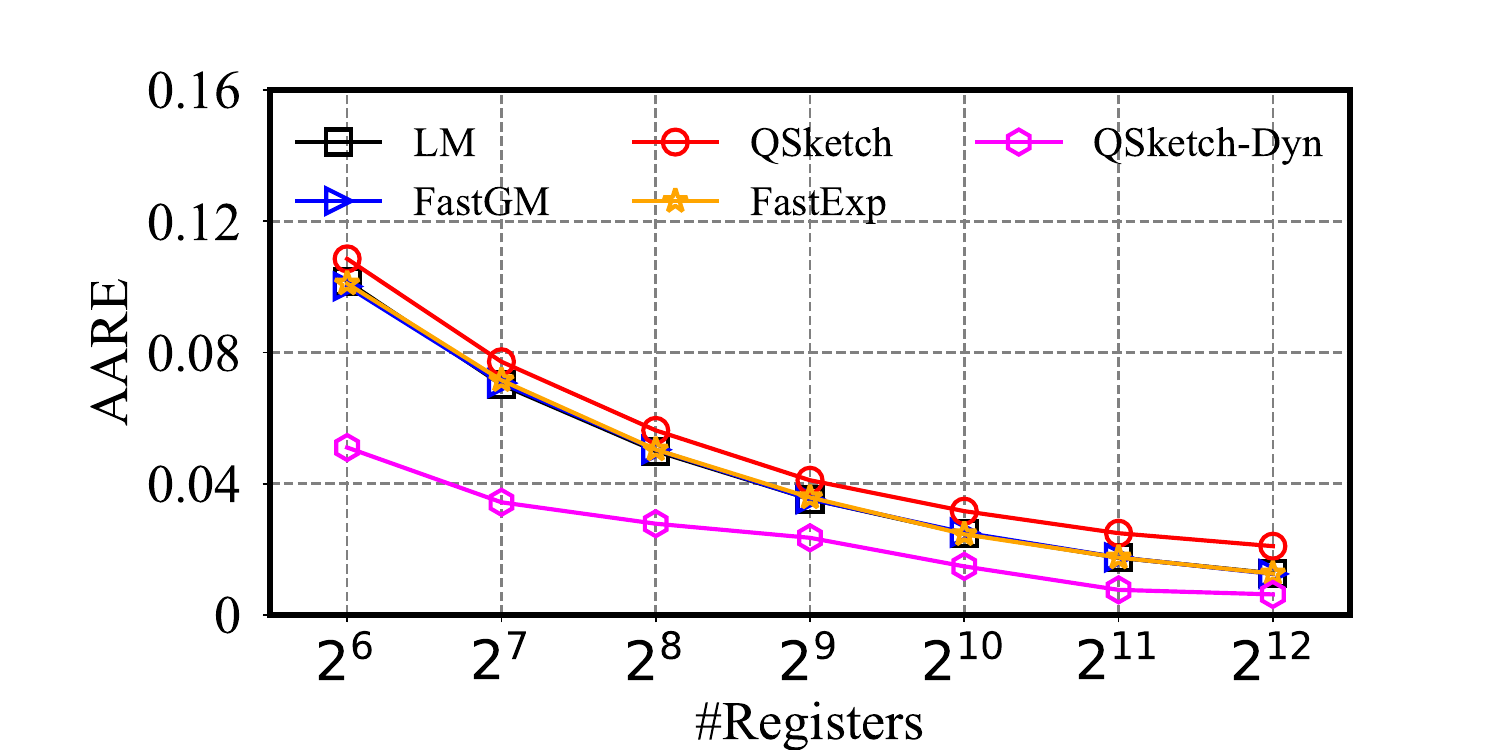}}
	\subfigure[Real-sim]{\includegraphics[width=0.329\textwidth, trim=40 5 60 35,clip]{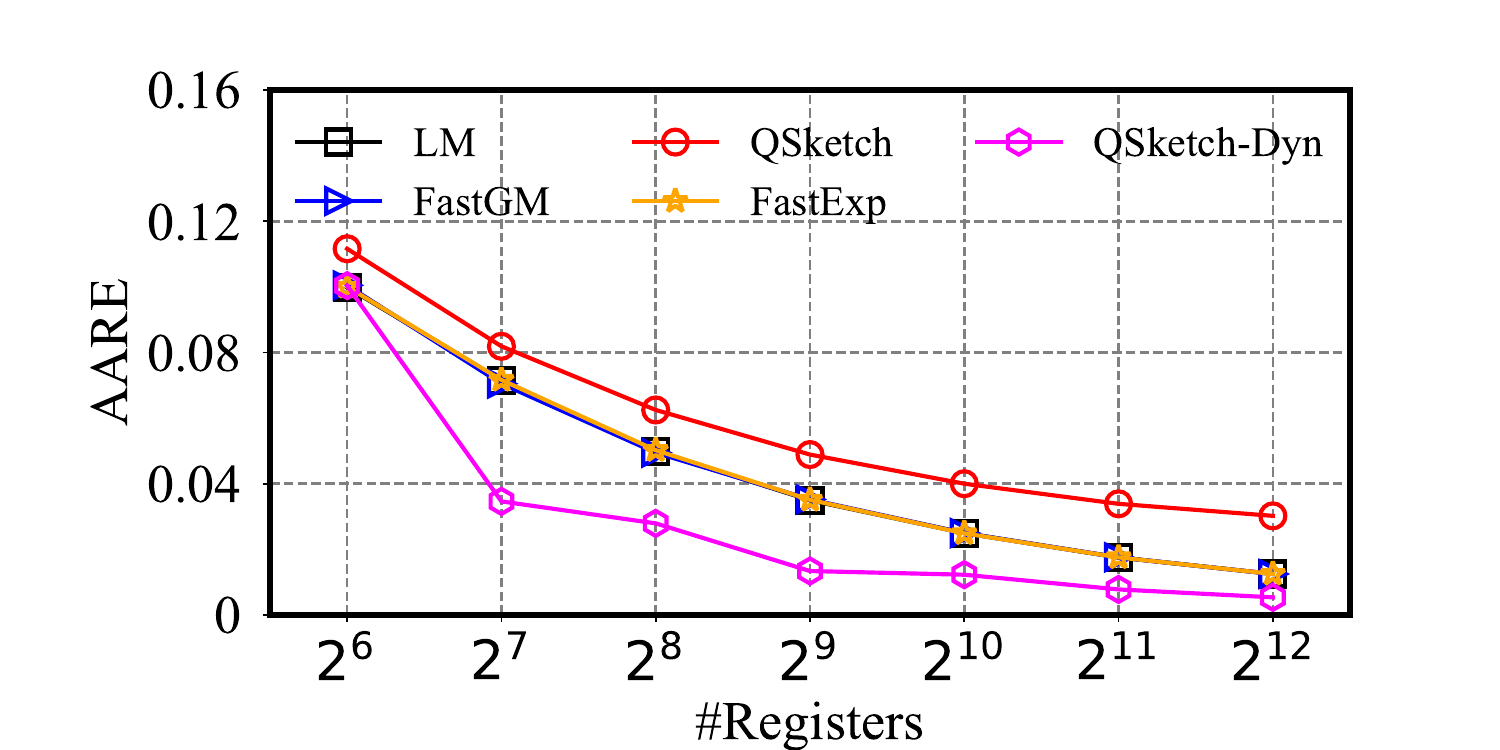}}
	\subfigure[Webspam]{\includegraphics[width=0.329\textwidth, trim=40 5 60 35,clip]{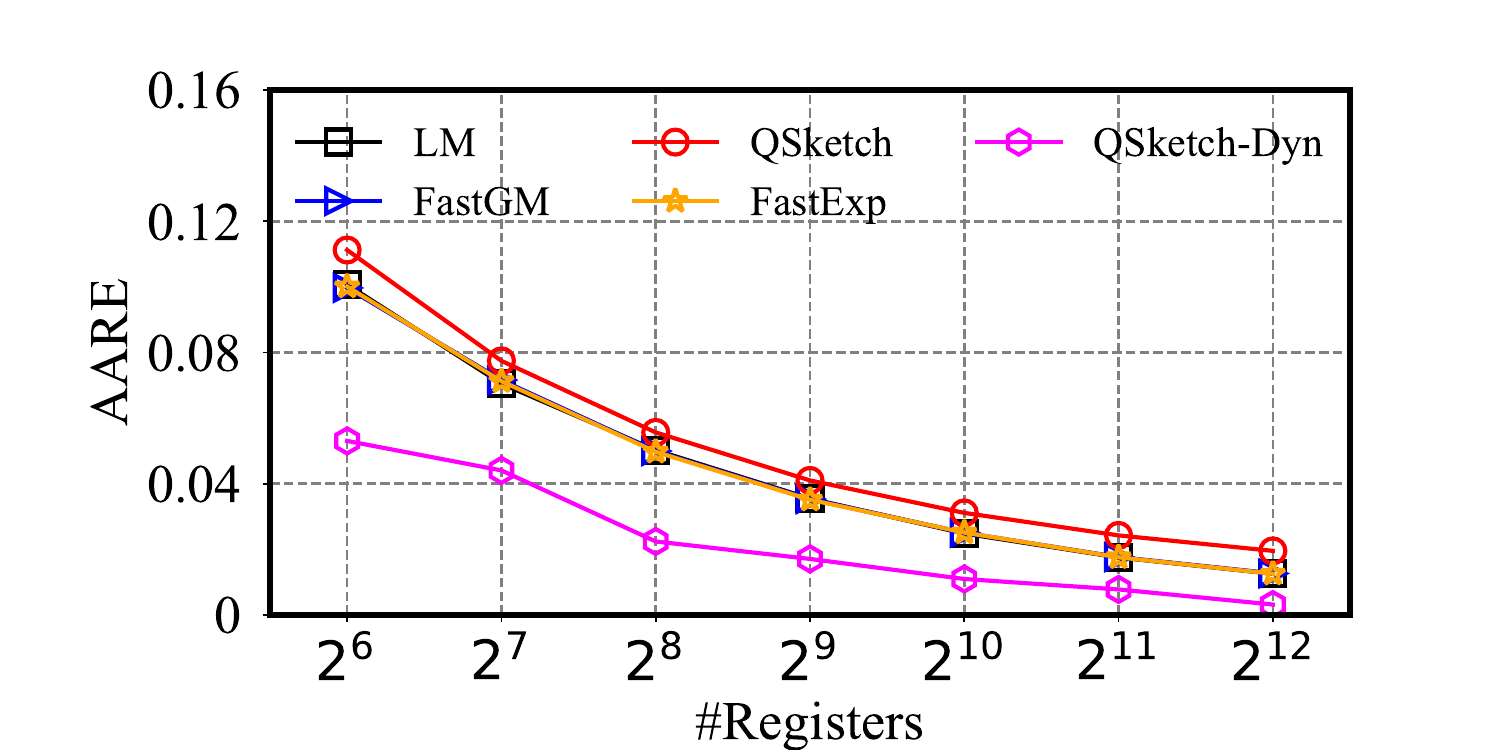}}
    \vspace{-5mm}
    \caption{Accuracy of all methods under different numbers of registers on real-world datasets.}
    \vspace{-4mm}
    \label{figs:realworld_acc}
\end{figure*}

\begin{figure*}[!t]
    \centering
	\subfigure[Gamma-10k]{\includegraphics[width=0.329\textwidth, trim=40 5 60 35,clip]{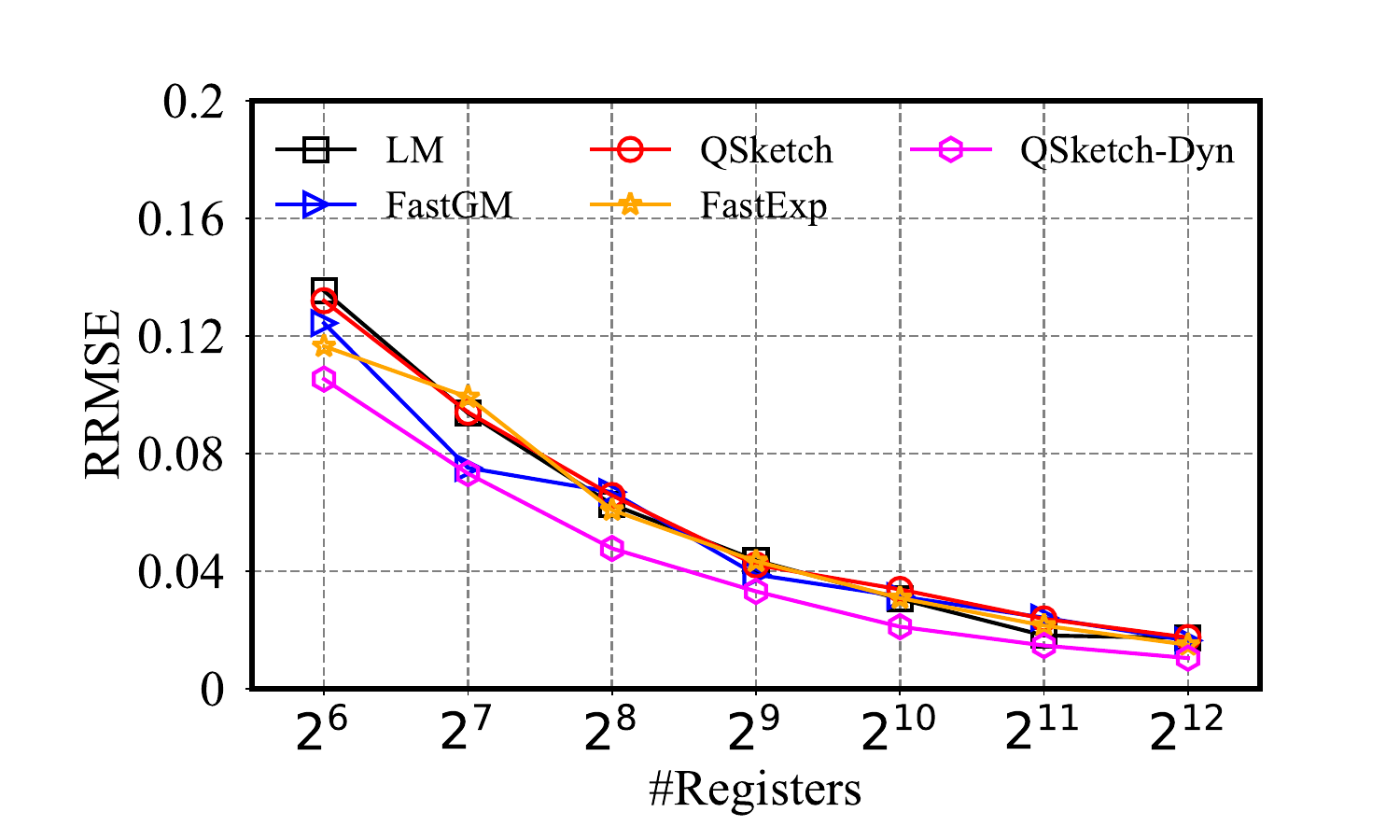}}    
	\subfigure[Gauss-10k]{\includegraphics[width=0.329\textwidth, trim=40 5 60 35,clip]{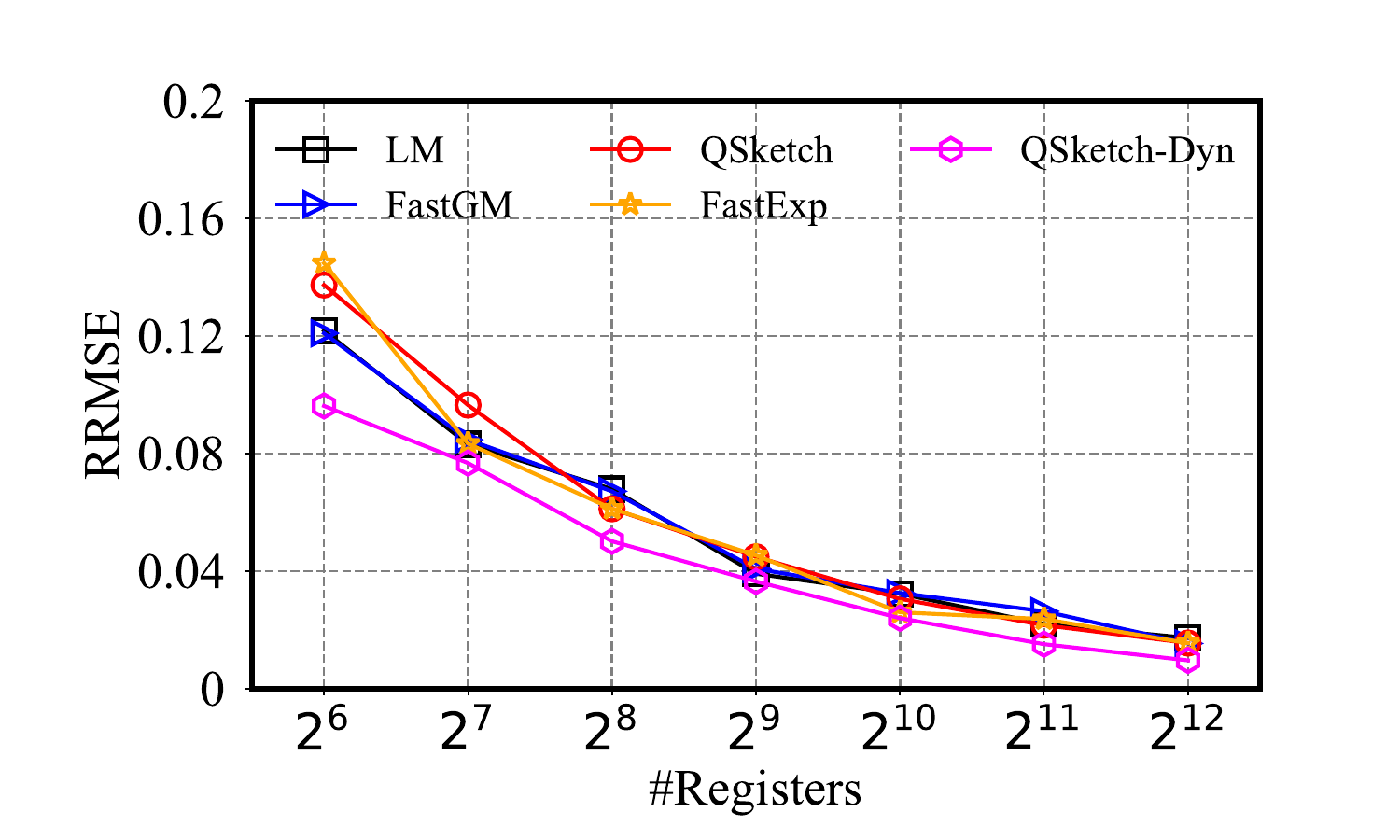}}
	\subfigure[Uniform-10k]{\includegraphics[width=0.329\textwidth, trim=40 5 60 35,clip]{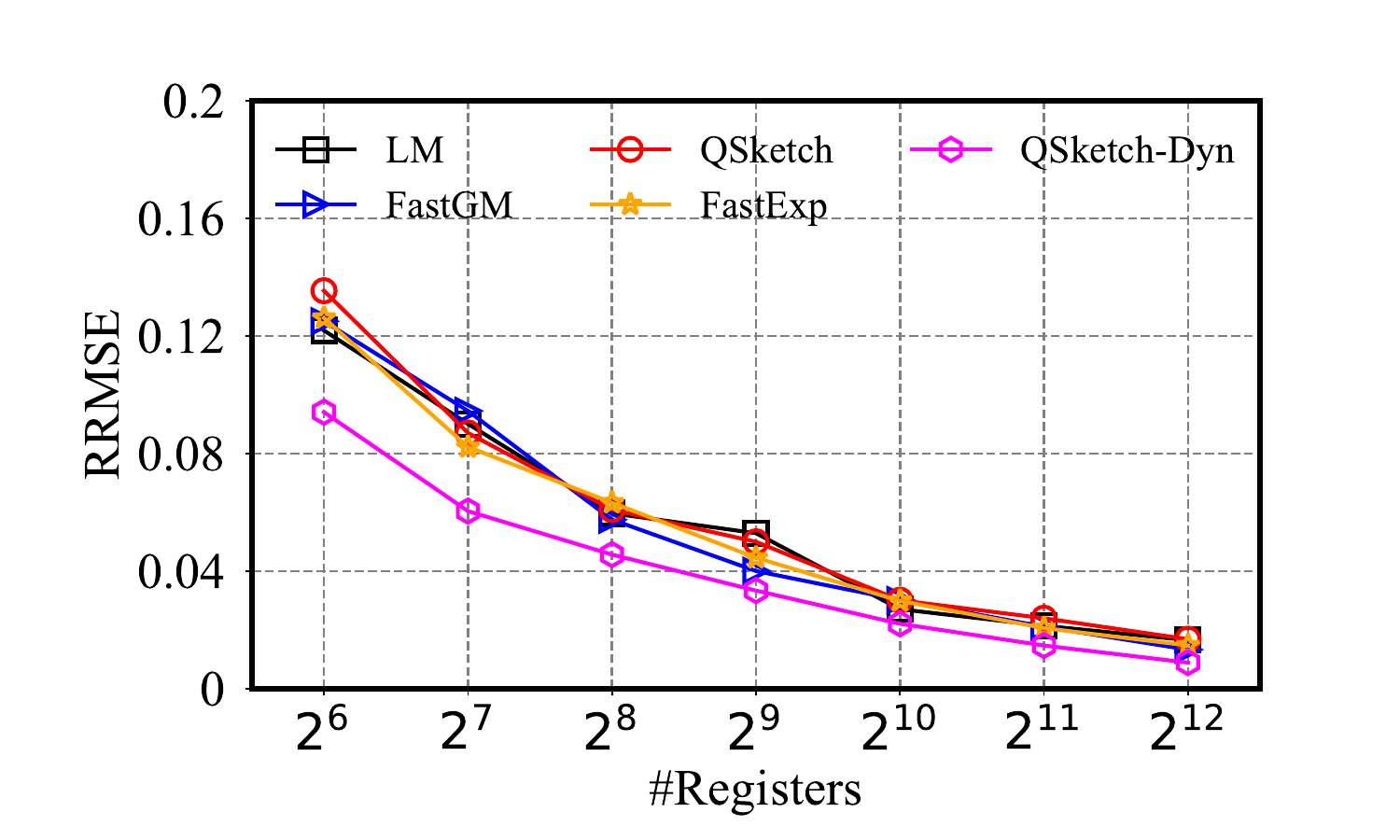}}
    \vspace{-5mm}
    \caption{Accuracy of all methods under different numbers of registers on synthetic datasets.}
    \label{figs:synthetic_acc}
    \vspace{-4mm}
\end{figure*}
\begin{figure*}[!t]
    \centering
	\subfigure[Gamma distribution]{\includegraphics[width=0.329\textwidth, trim=40 5 60 35,clip]{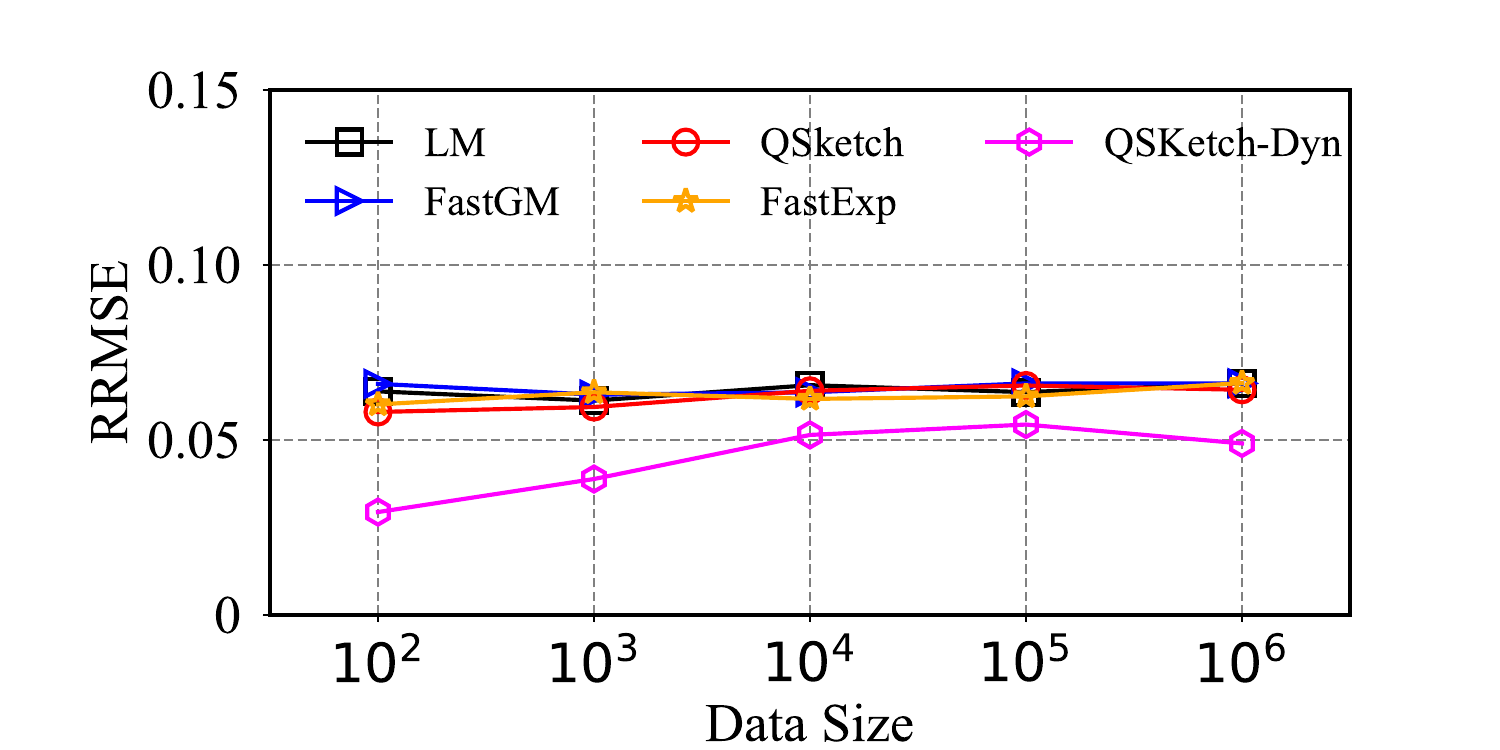}}    
	\subfigure[Gauss distribution]{\includegraphics[width=0.329\textwidth, trim=40 5 60 35,clip]{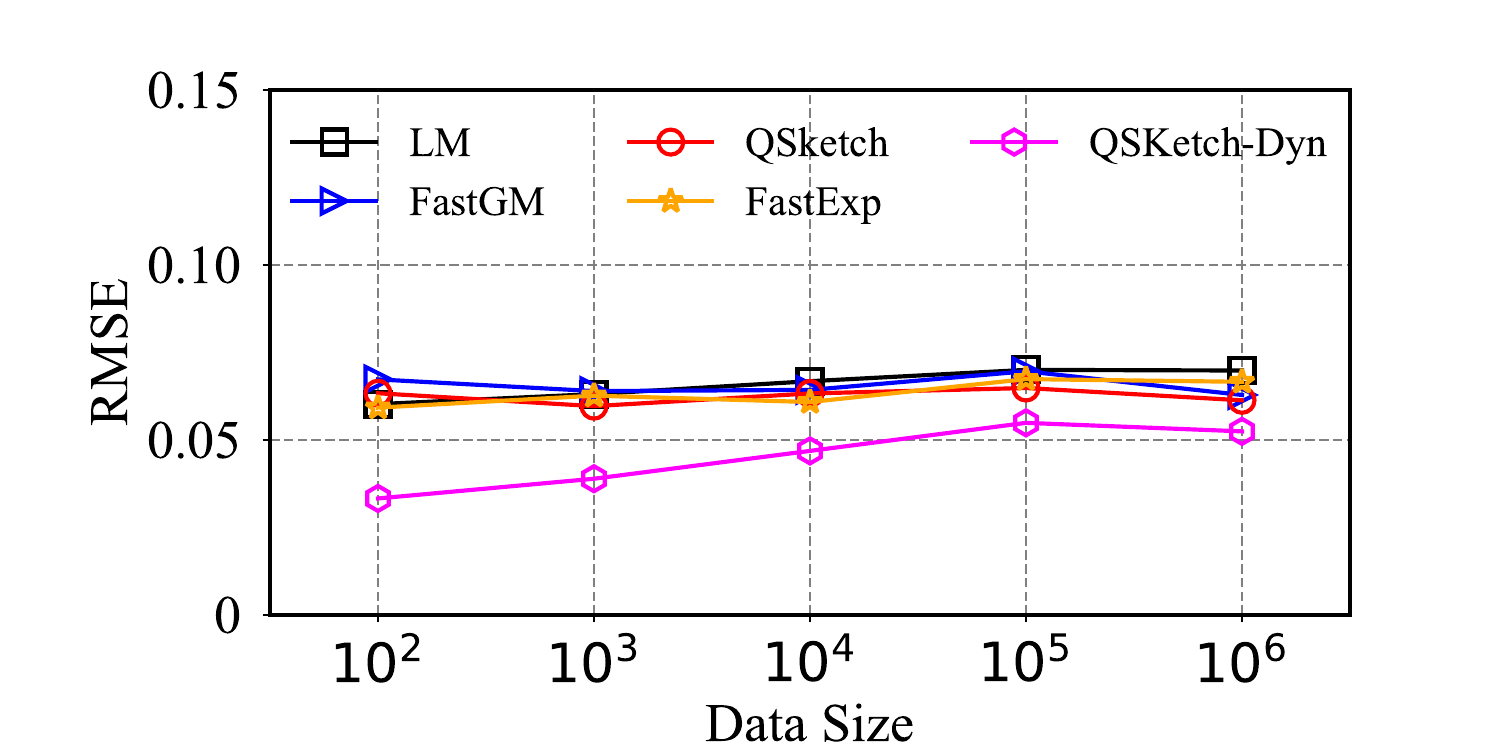}}
	\subfigure[Uniform distribution]{\includegraphics[width=0.329\textwidth, trim=40 5 60 35,clip]{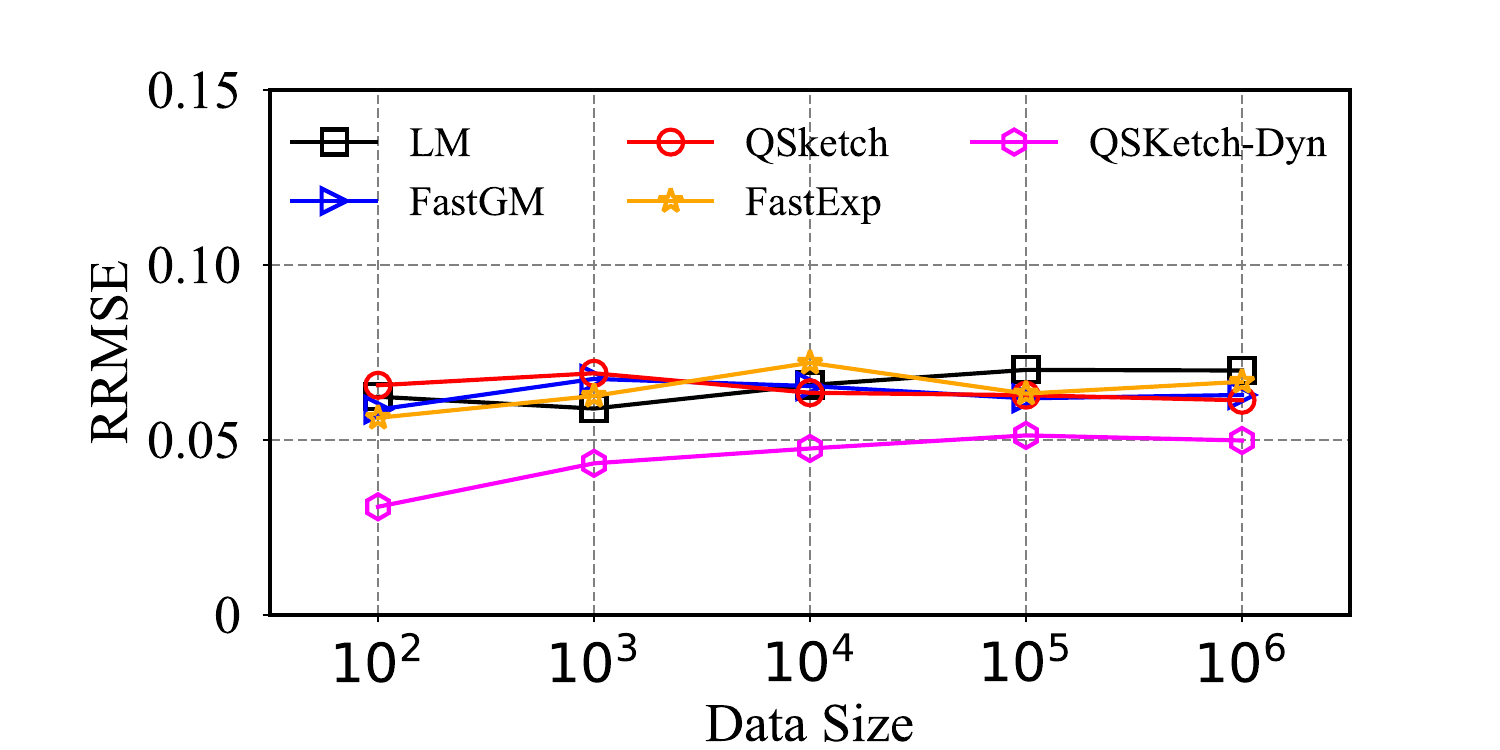}}
    \vspace{-5mm}
    \caption{Accuracy of all methods under different data sizes on synthetic datasets.}
    \label{figs:data_size}
\end{figure*}

\subsection{Baselines}
We compare QSketch and QSketch-Dyn, with state-of-the-art methods,
Lemiesz's method~\cite{lemiesz2021algebra} (represented as \textbf{LM}), FastGM~\cite{zhang2023fast} and FastExp Sketch~\cite{lemiesz2023efficient}.
All baseline methods maintain a sketch with $m$ 64-bit registers.
QSketch and QSketch-Dyn are truncated with $r_\text{min}=-127$ and $r_\text{max}=127$,
i.e., they all use 8-bit integer registers by default.
We assign each algorithm the same number of registers,
which means QSketch and QSketch-Dyn use about 1/8 of the memory space of baseline methods.
Following~\cite{hlll}, we use a 32-bit word to hold multiple short-bit registers.
For example, with each register set to 8 bits, a 32-bit word can hold $\lfloor \frac{32}{8} \rfloor = 4$ registers.

\subsection{Metrics}
\noindent $\bullet$ \textbf{Accuracy.}
We use \textit{Relative Root Mean Square Error} (RRMSE) and \textit{Average Absolute Relative Error} (AARE) to evaluate the estimation accuracy
on single-stream datasets and multi-stream datasets, respectively.
In detail, the RRMSE of estimation $\hat{C}$ is defined as
\[
RRMSE(\hat{C}) = \frac{\sqrt{\mathbb{E}[(\hat{C} - C)^2]}}{C},
\]
and the AARE is defined as
\[
AARE =\frac{1}{n} \sum_{i=1}^n \frac{|\hat{C}_i - C_i|}{|C_i|}.
\]

\noindent $\bullet$ \textbf{Efficiency.}
We use \textbf{Throughput} (Million updates per second, Mops) to evaluate the update speed for incoming elements,
and \textbf{Esimation time} to assess the time taken to calculate the weighted cardinality from the sketch.
All experimental results are empirically computed from 100 independent runs by default.

\subsection{Accuracy Analysis}
\subsubsection{Results on Real-World Datasets}
Figure~\ref{figs:realworld_acc} shows the results on accuracy concerning the number of registers in a sketch on real-world datasets.
Specially, we vary the number of registers in each sketch $m\in\{2^6, 2^7, 2^8, 2^9, 2^{10}, 2^{11}, 2^{12}\}$.
For dataset Twitter and the other 3 document datasets, we evaluate RRMSE and AARE w.r.t. the number of registers, respectively.
Notably, QSketch demonstrates comparable performance to other baseline methods across all datasets. 
Conversely, QSketch-Dyn outperforms its competitors, leveraging the dynamic nature of the sketch.
For example, on the dataset Twitter, our method QSketch-Dyn is $30\%$ more accurate than alternative methods with the number of registers $m = 2^8$. 
It is imperative to emphasize that QSketch and QSketch-Dyn utilize only $1/8$ of the memory compared to LM and FastGM.

\subsubsection{Results on Synthetic Datasets}
To comprehensively assess the efficacy of QSketch across diverse scenarios, 
we conduct a thorough evaluation comparing its performance with that of other baseline methods. 
This evaluation encompasses a range of factors including data distribution, dataset scale, register count, and register size.

\noindent \textbf{Performance under different data distribution.} 
We compare our methods QSketch and QSketch-Dyn with other methods on synthetic datasets from different distributions.
Figure \ref{figs:synthetic_acc} illustrates the comparative performance. 
Remarkably, QSketch-Dyn consistently outperforms other methods across all distributions, the same as real-world dataset results.

\noindent \textbf{Performance under different dataset sizes.}
Next, we explore the performance of our methodologies across varying dataset sizes. 
We generate datasets from three distributions at different scales ranging from $10^2$ to $10^6$. 
The results of the remaining two distributions are summarized in the Appendix.
The number of registers for all methods is fixed at $2^8$.
As shown in Figure~\ref{figs:data_size}, 
QSketch, LM, FastGM, and FastExp Sketch estimation errors remain consistent across all dataset scales. 
However, the performance of QSketch-Dyn shows a slight improvement with increasing dataset scale, with the estimation error stabilizing around $0.05$ for dataset sizes exceeding $10^4$. 
This phenomenon primarily stems from the fact that, with smaller dataset sizes, most registers in QSketch-Dyn are populated by only one element, resulting in nearly exact counting.

\noindent \textbf{Performance under different register size.}
\begin{figure*}[!t]
    \centering
	\subfigure[QSketch]{\includegraphics[width=0.429\textwidth, trim=40 5 60 35,clip]{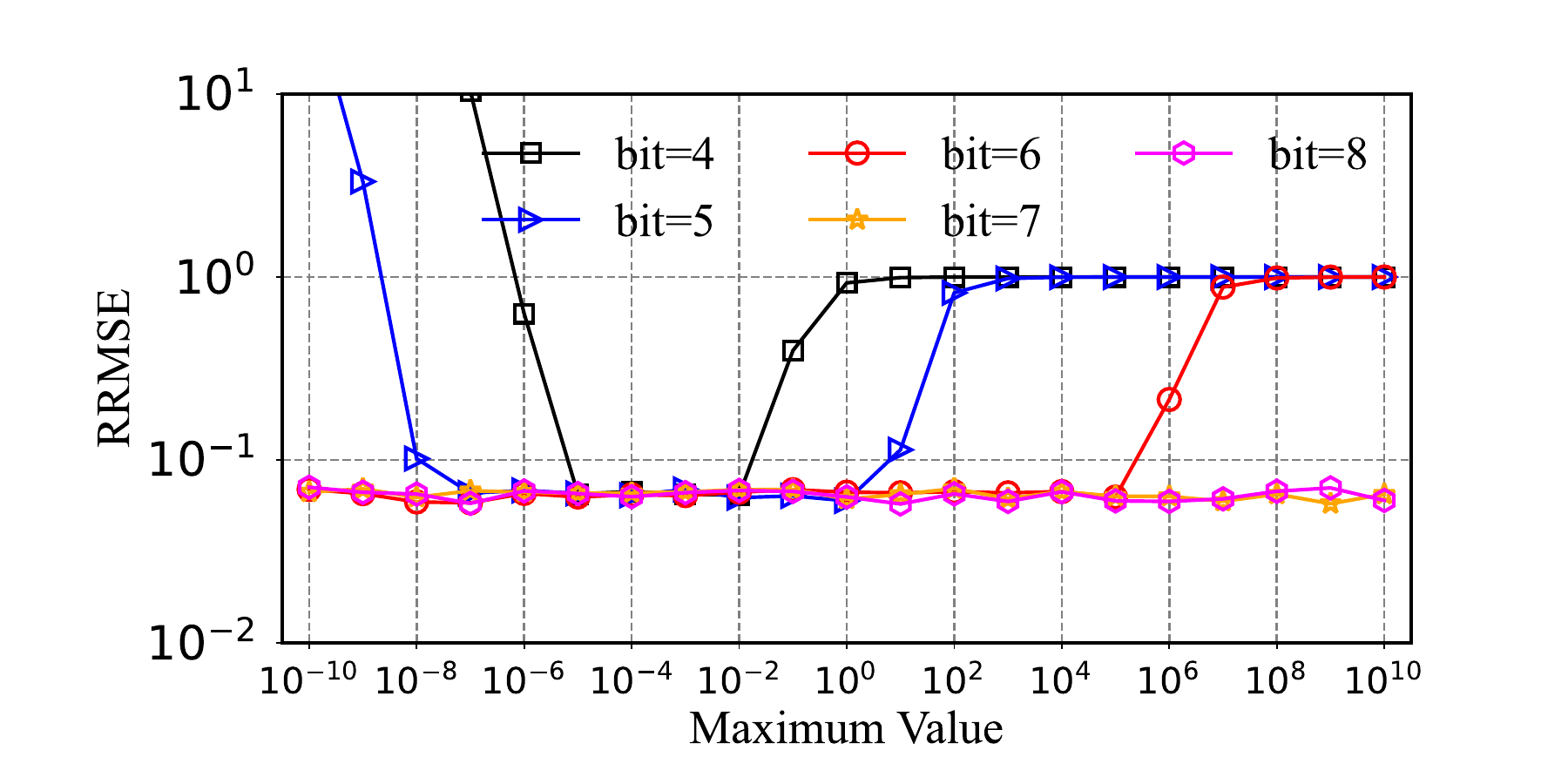}}  
	\subfigure[QSketch-dyn]{\includegraphics[width=0.429\textwidth, trim=40 5 60 35,clip]{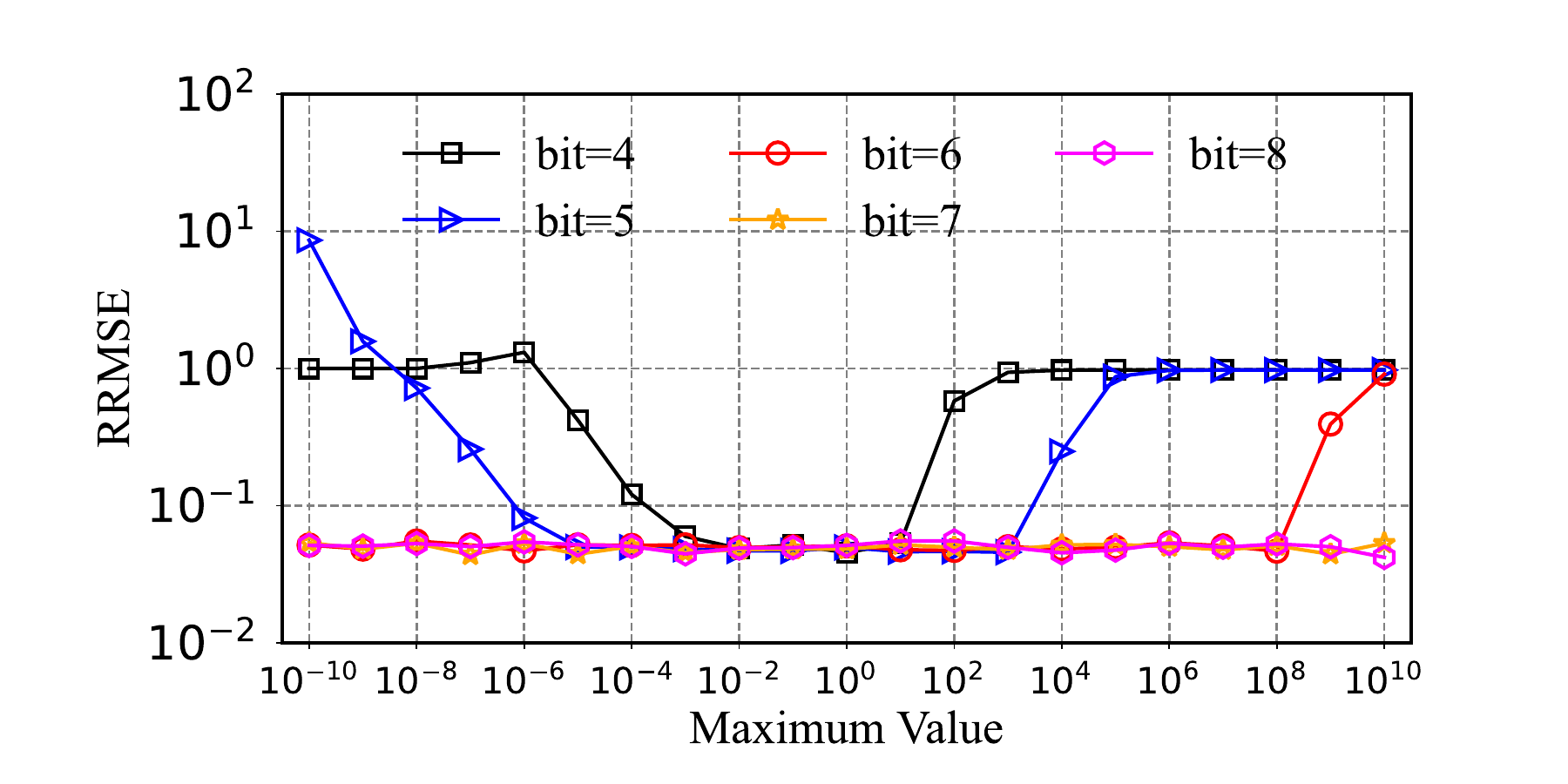}}
    \vspace{-5mm}
    \caption{Accuracy of our methods QSketch and QSketch-Dyn under different register sizes on synthetic datasets.}
    \label{figs:register_size}
    \vspace{-4mm}
\end{figure*}
As mentioned in Theorem~\ref{theorem1}, the bit size of the sketch's registers also influences its performance. 
Figure~\ref{figs:register_size} illustrates the estimation error of QSketch and QSketch-Dyn on the Uniform-10k distribution, considering the maximum value of the distribution ranging from $10^{-10}$ to $10^{10}$ (i.e., weighted cardinality ranging from $5\times 10^{-7}$ to $5 \times 10^{13}$).
The number of registers of both methods is set to $2^8$.
It is evident that when employing 4- or 5-bit registers, both QSketch and QSketch-Dyn offer accurate estimations within a limited range. 
However, with a bit size increase to 7 or 8, both methods consistently perform well across all values, aligning with the findings of Theorem~\ref{theorem1}.

\subsection{Efficiency Analysis}
For efficiency, we measure the \textbf{Throughput} (Millions of updates per second) as the update speed of a sketch for incoming elements, 
and the \textbf{Estimation time} as the time taken to calculate the weighted cardinality from the sketch.

\begin{figure*}[!t]
    \centering
    \subfigure[Twitter]{\includegraphics[width=0.329\textwidth, trim=20 0 60 20,clip]{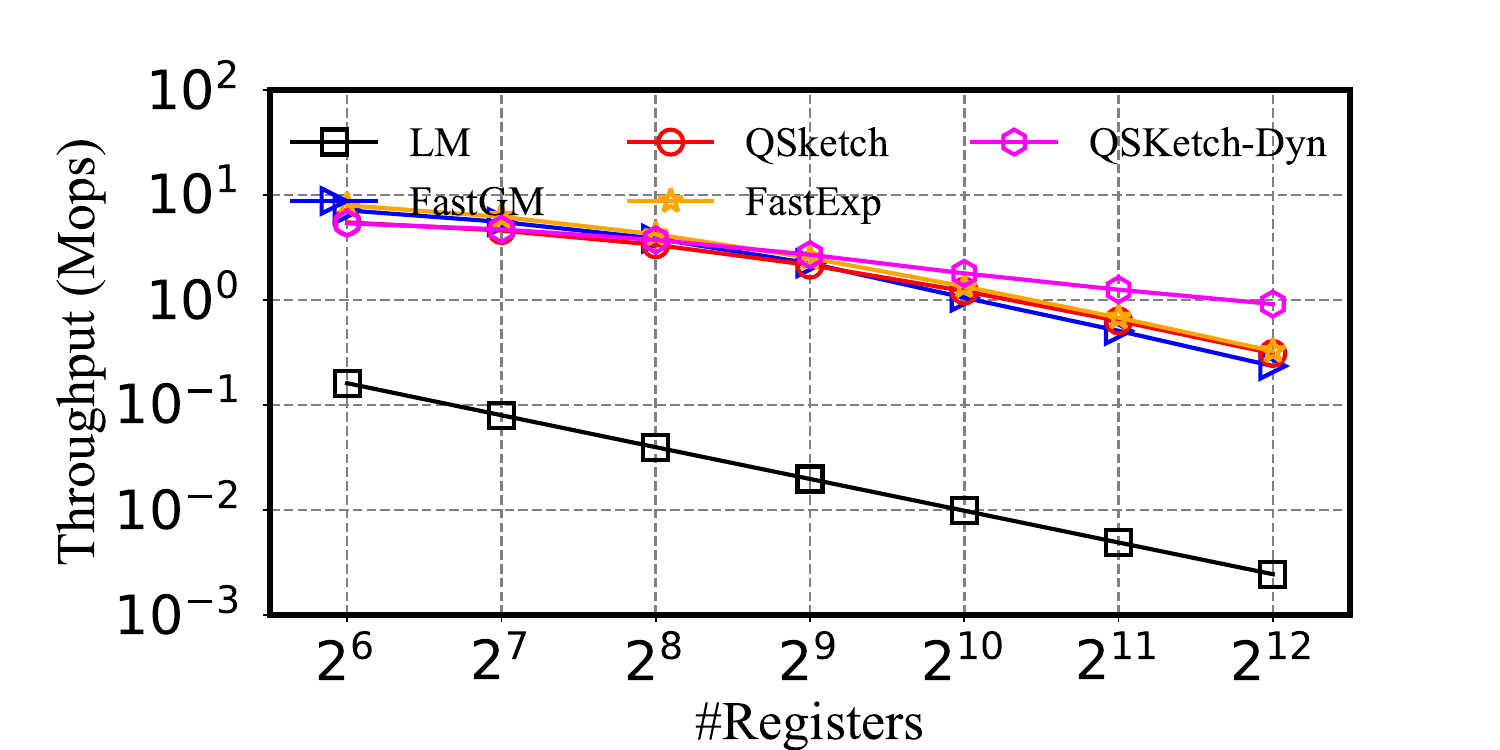}}
    \subfigure[Libimseti]{\includegraphics[width=0.329\textwidth, trim=20 0 60 20,clip]{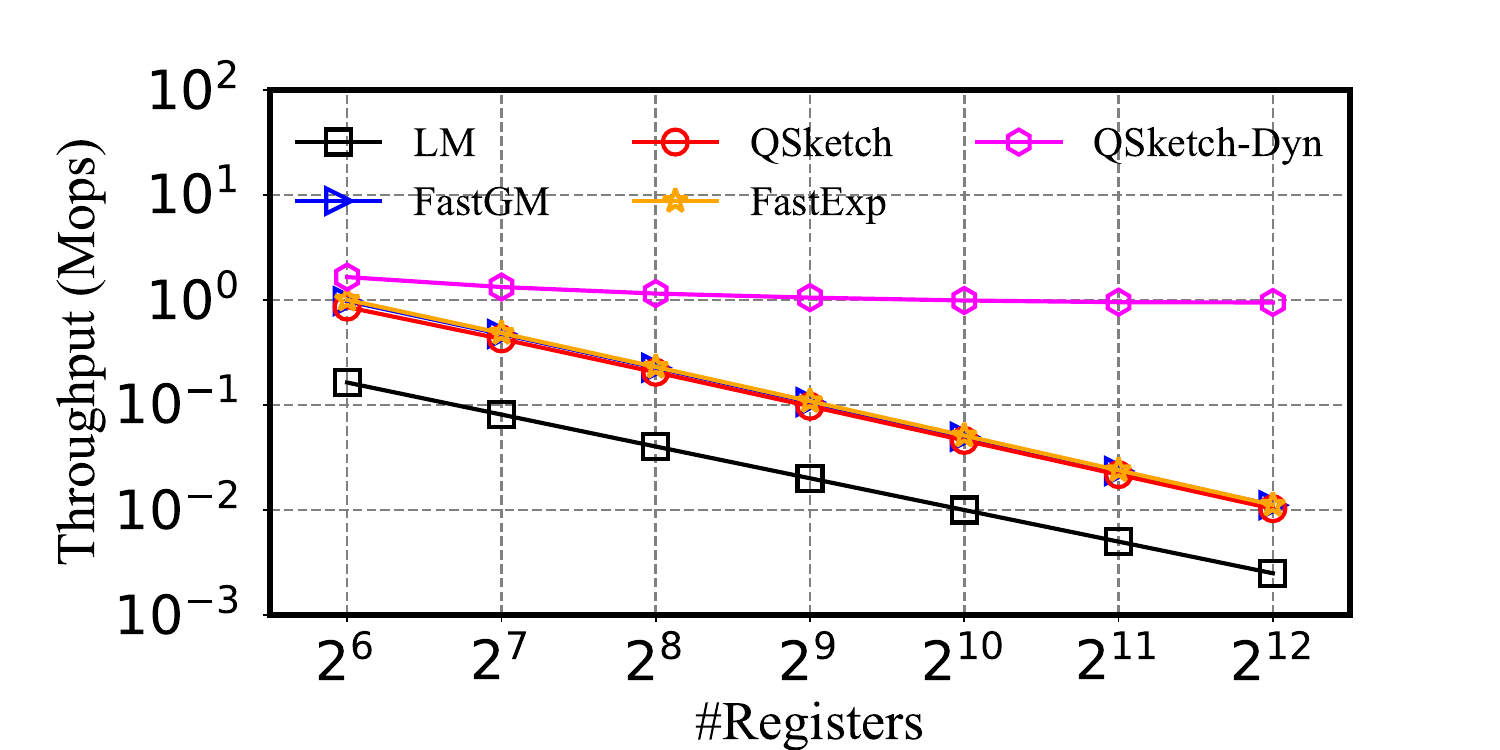}}
    \subfigure[News20]{\includegraphics[width=0.329\textwidth, trim=20 0 60 20,clip]{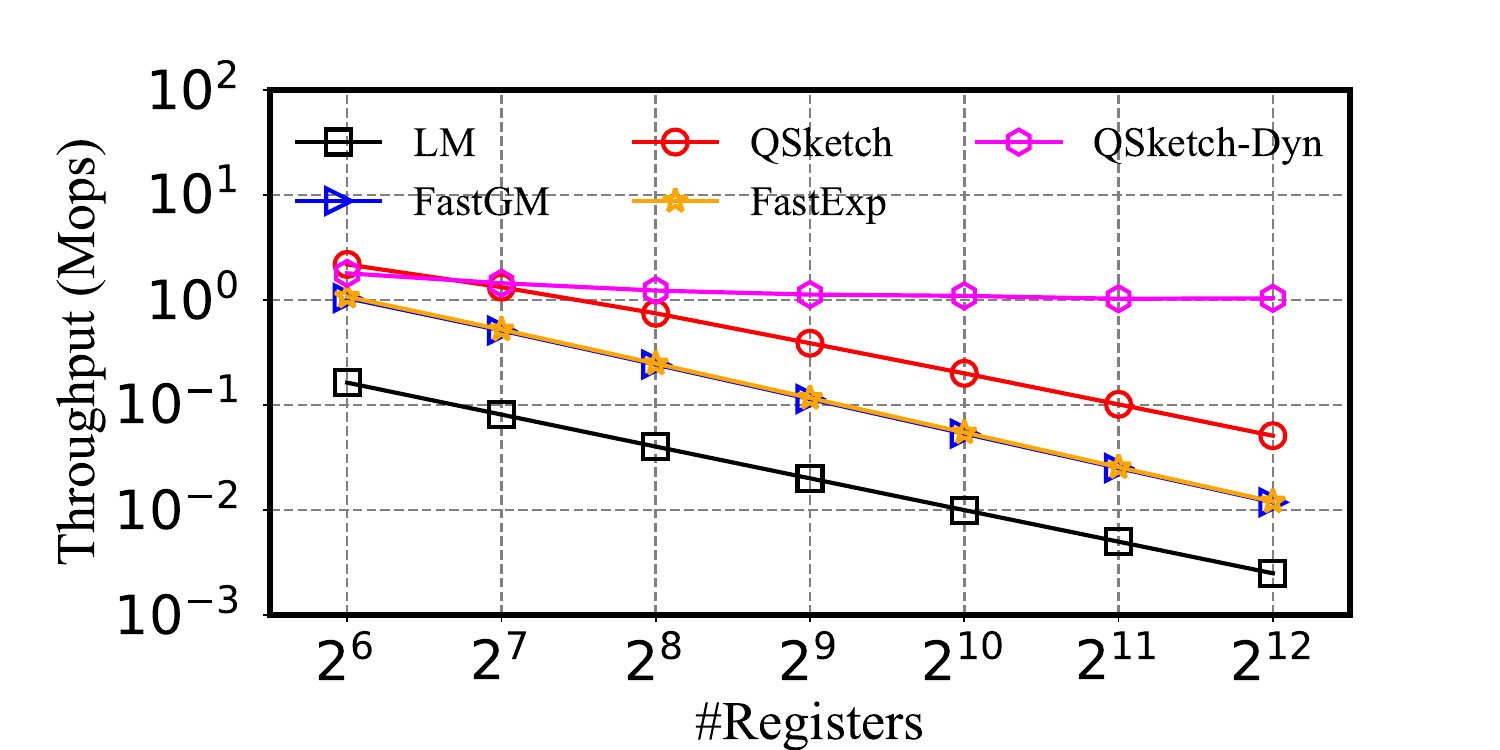}}
    \subfigure[Rcv1]{\includegraphics[width=0.329\textwidth, trim=20 0 60 20,clip]{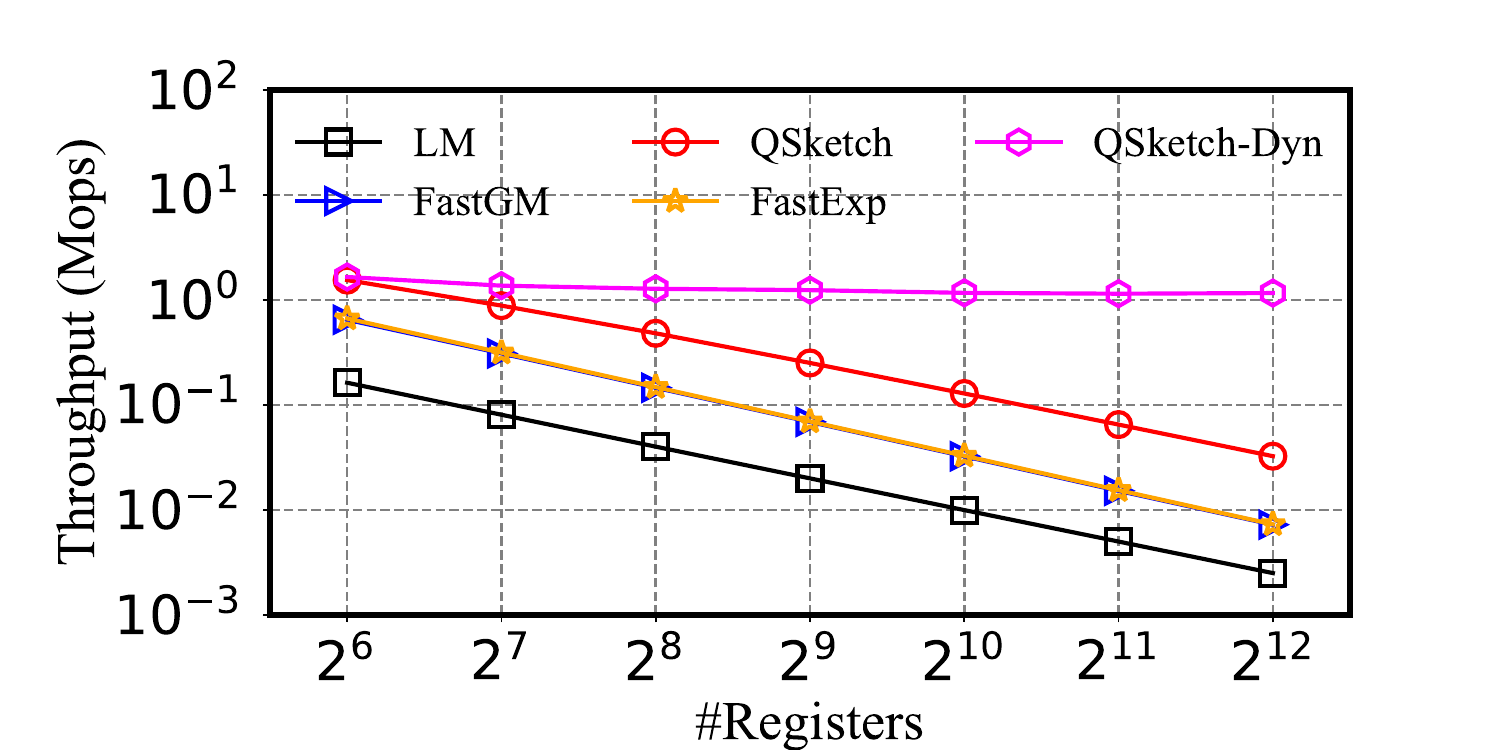}}
	\subfigure[Real-sim]{\includegraphics[width=0.329\textwidth, trim=20 0 60 20,clip]{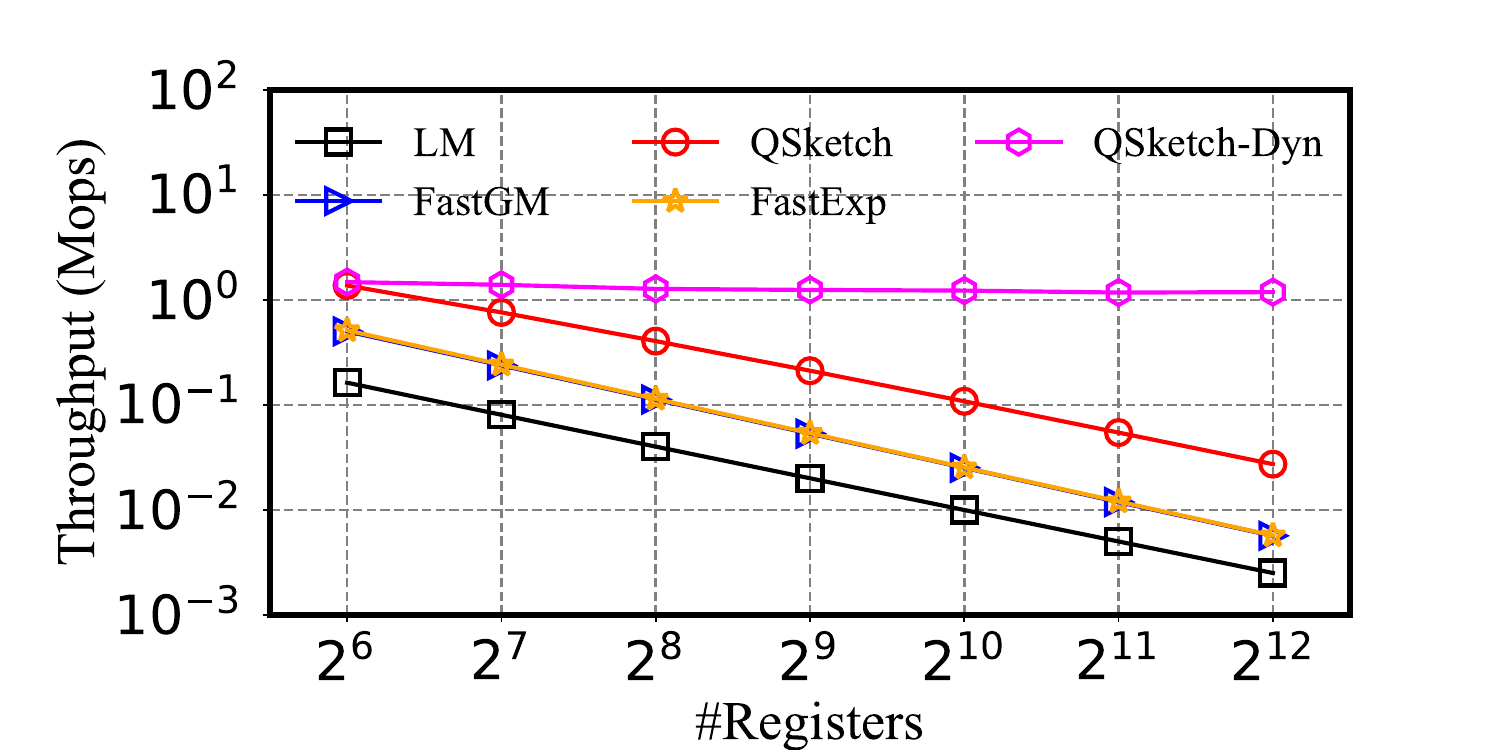}}
	\subfigure[Webspam]{\includegraphics[width=0.329\textwidth, trim=20 0 60 20,clip]{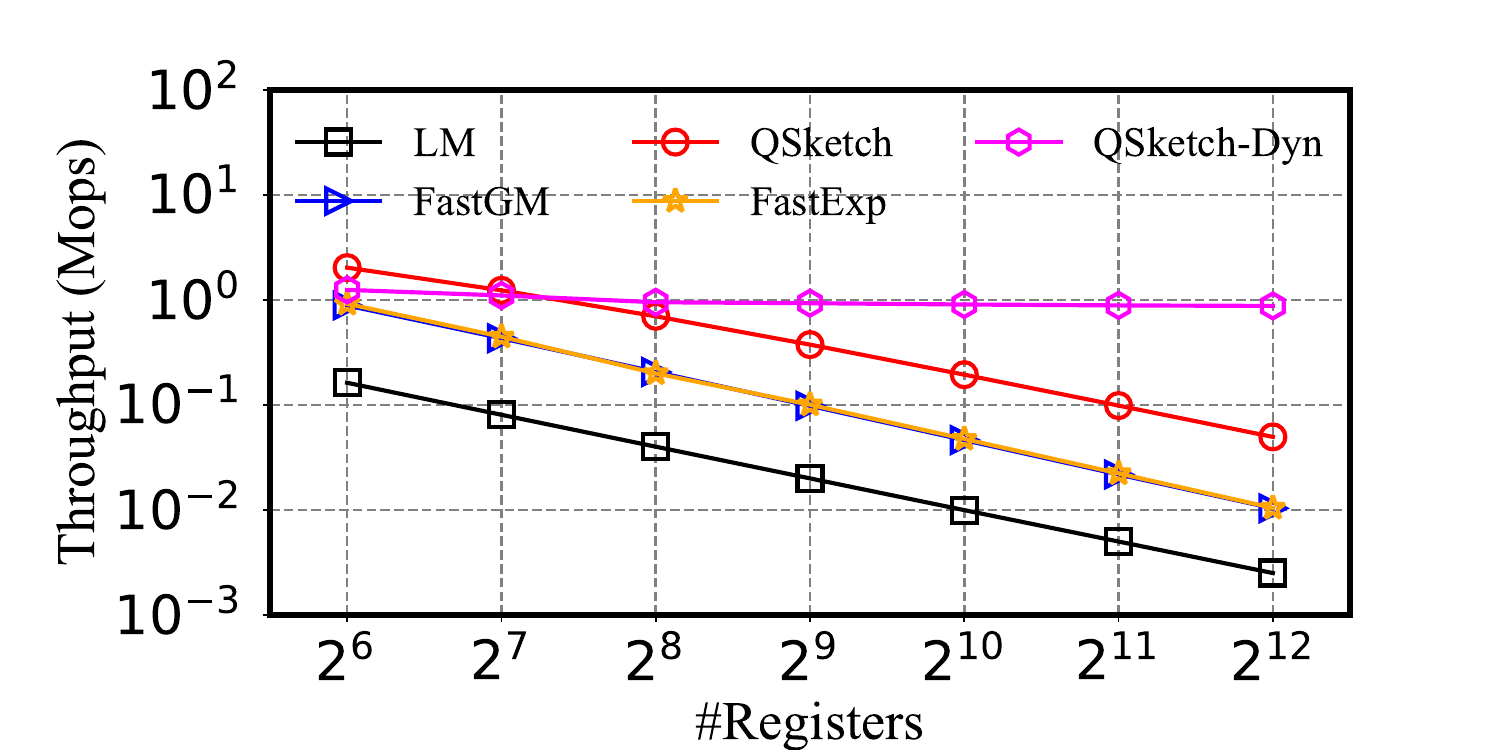}}
    \vspace{-5mm}
    \caption{Update throughput of all methods under different numbers of registers on real-world datasets.}
    \vspace{-4mm}
    \label{figs:realworld_update_time}
\end{figure*}

\subsubsection{Results on Real-World Datasets}
Figure~\ref{figs:realworld_update_time} shows the results of the update throughput on real-world datasets.
The throughput for LM, FastGM, FastExp Sketch, and QSketch demonstrates a decrease with more registers in the sketch. 
Moreover, the update throughput for FastGM and FastExp Sketch exhibits similarity and is faster than LM.
For QSketch, since it uses packed integers for implementation, it needs fewer memory accesses than FastGM and FastExp,
which leads to a large update throughput on most datasets.
The update throughput for QSketch-Dyn remains nearly consistent across varying numbers of registers. 
Specifically, on dataset Rcv1, the update time for QSketch-Dyn is approximately 2 to 3 orders of magnitude shorter compared to FastGM and LM, respectively.

\subsubsection{Results on Synthetic Datasets}

\begin{figure*}[!t]
    \centering
    \subfigure[Gamma-10k]{\includegraphics[width=0.329\textwidth,trim=20 5 60 30,clip]{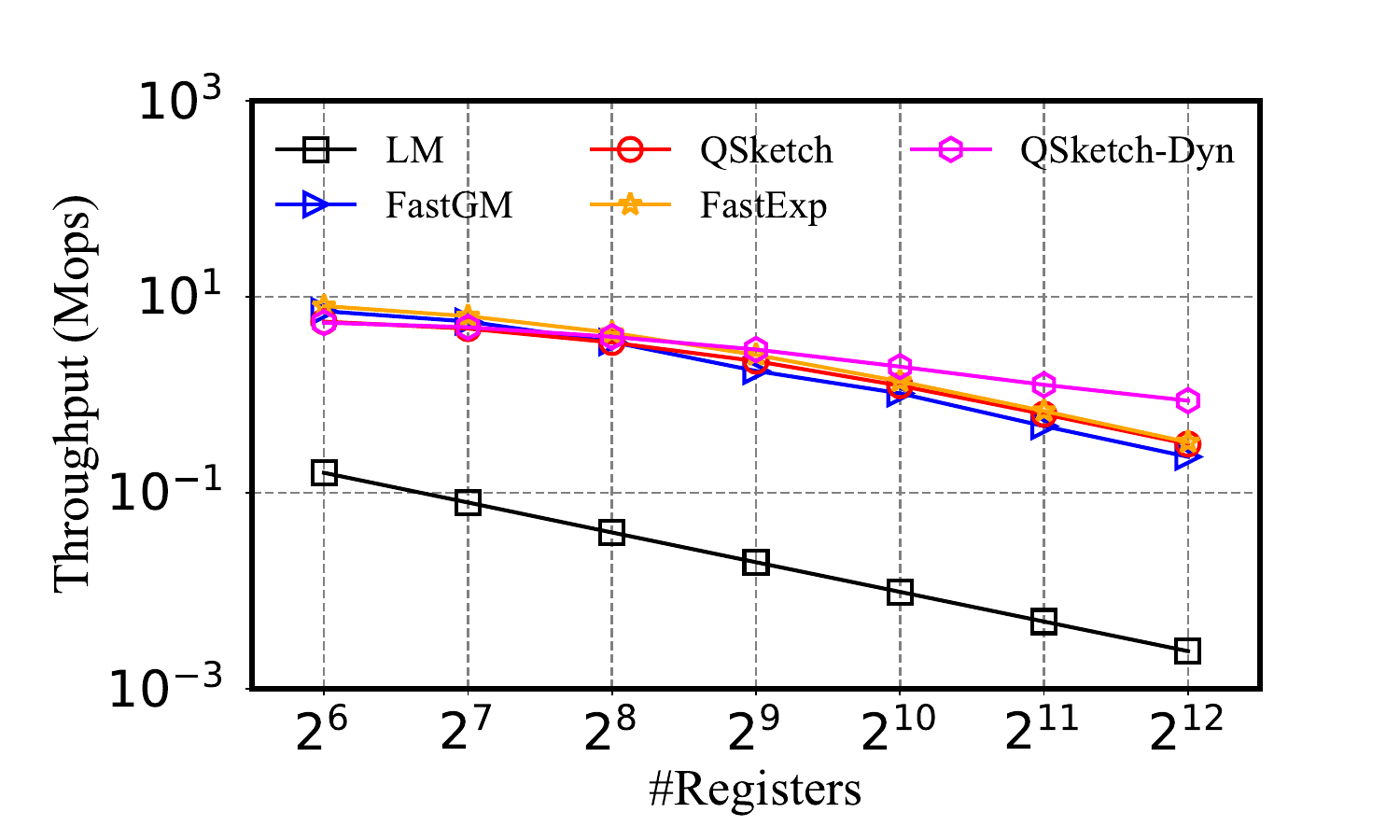}}
	\subfigure[Gauss-10k]{\includegraphics[width=0.329\textwidth,trim=20 5 60 30,clip]{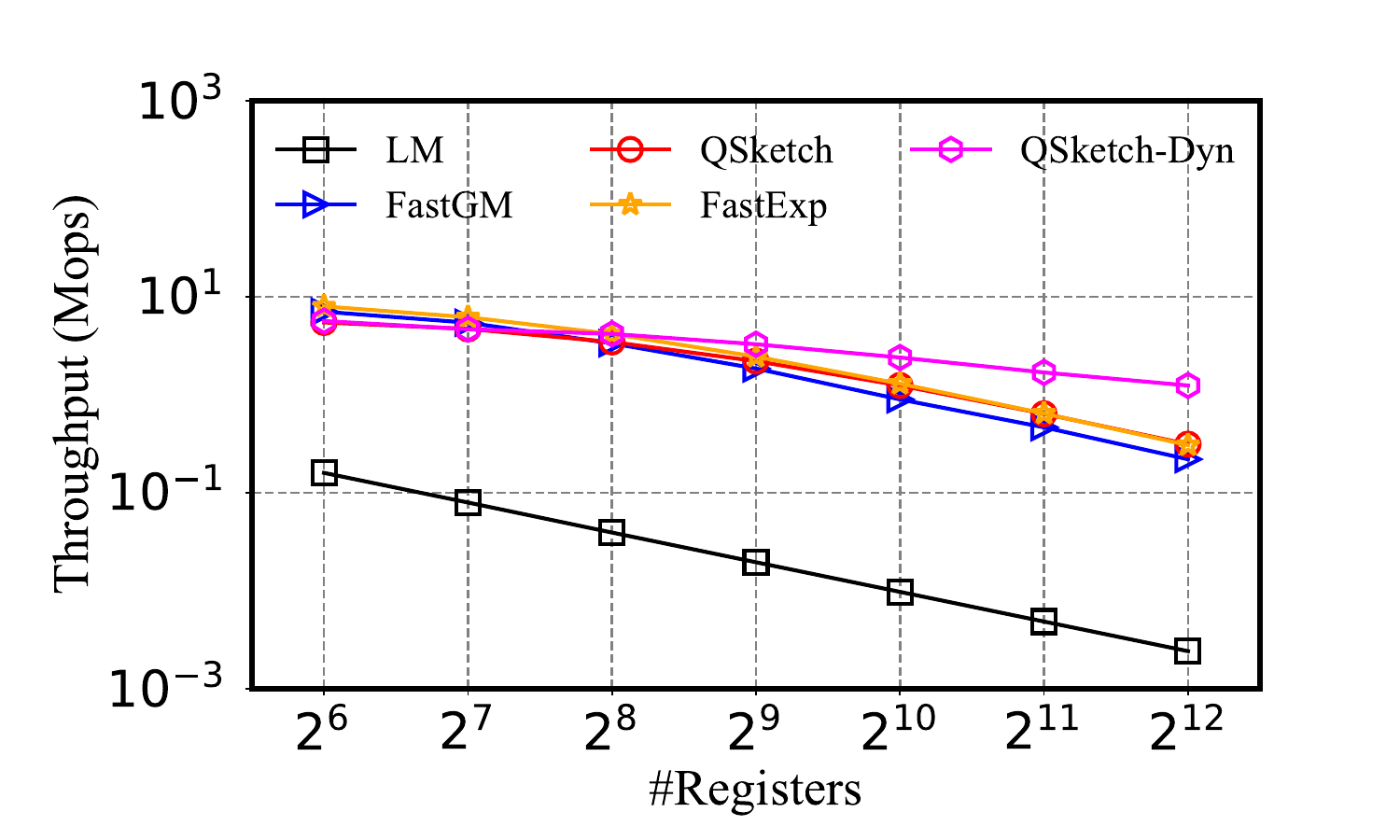}}
	\subfigure[Uniform-10k]{\includegraphics[width=0.329\textwidth,trim=20 5 60 30,clip]{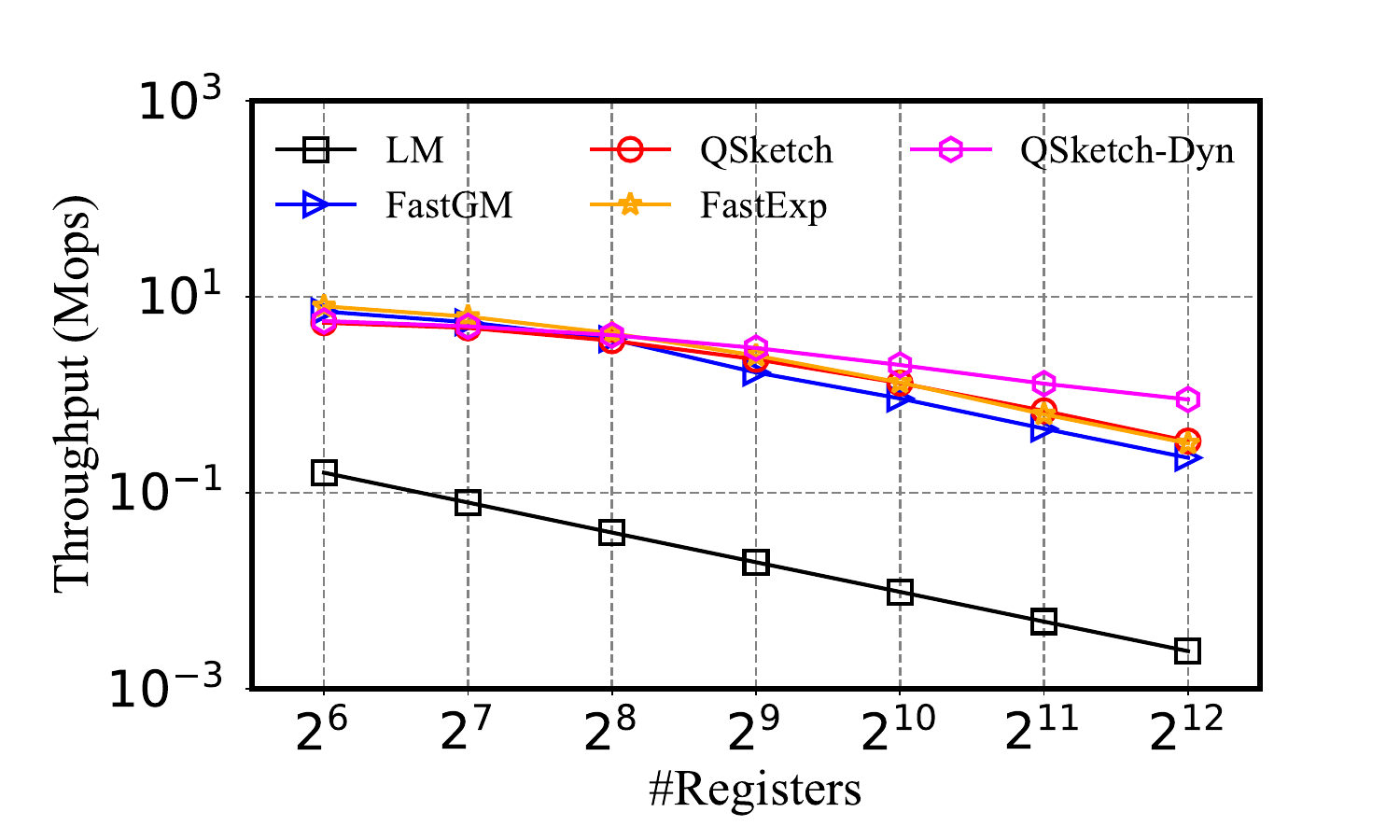}}
    \vspace{-5mm}
    \caption{Update throughput of all methods under different numbers of registers on synthetic datasets.}
    \vspace{-3mm}
    \label{figs:synthetic_update_time}
\end{figure*}

\begin{figure*}[t]
    \centering
    \subfigure[Gamma-10k]{\includegraphics[width=0.329\textwidth,trim=26 5 60 30,clip]{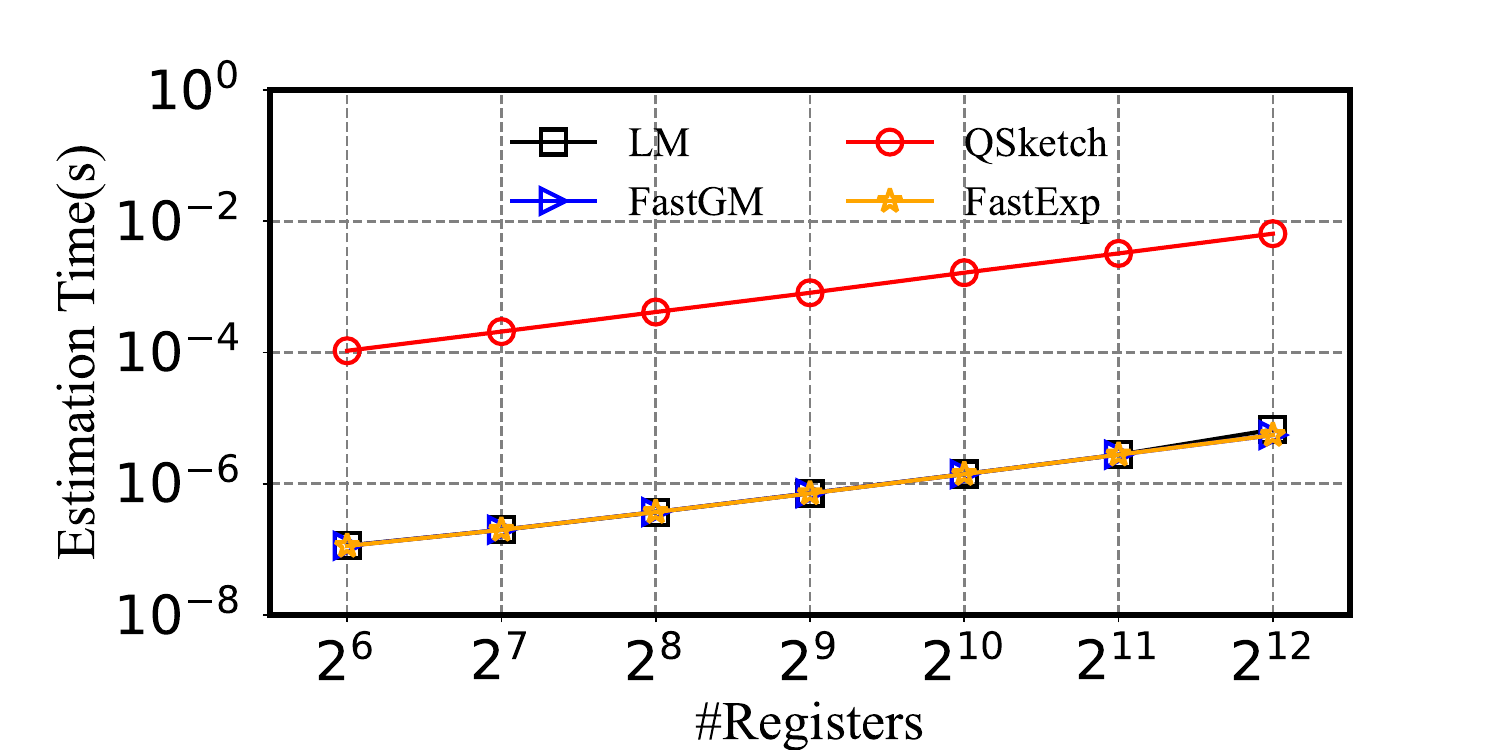}}
	\subfigure[Gauss-10k]{\includegraphics[width=0.329\textwidth,trim=26 5 60 30,clip]{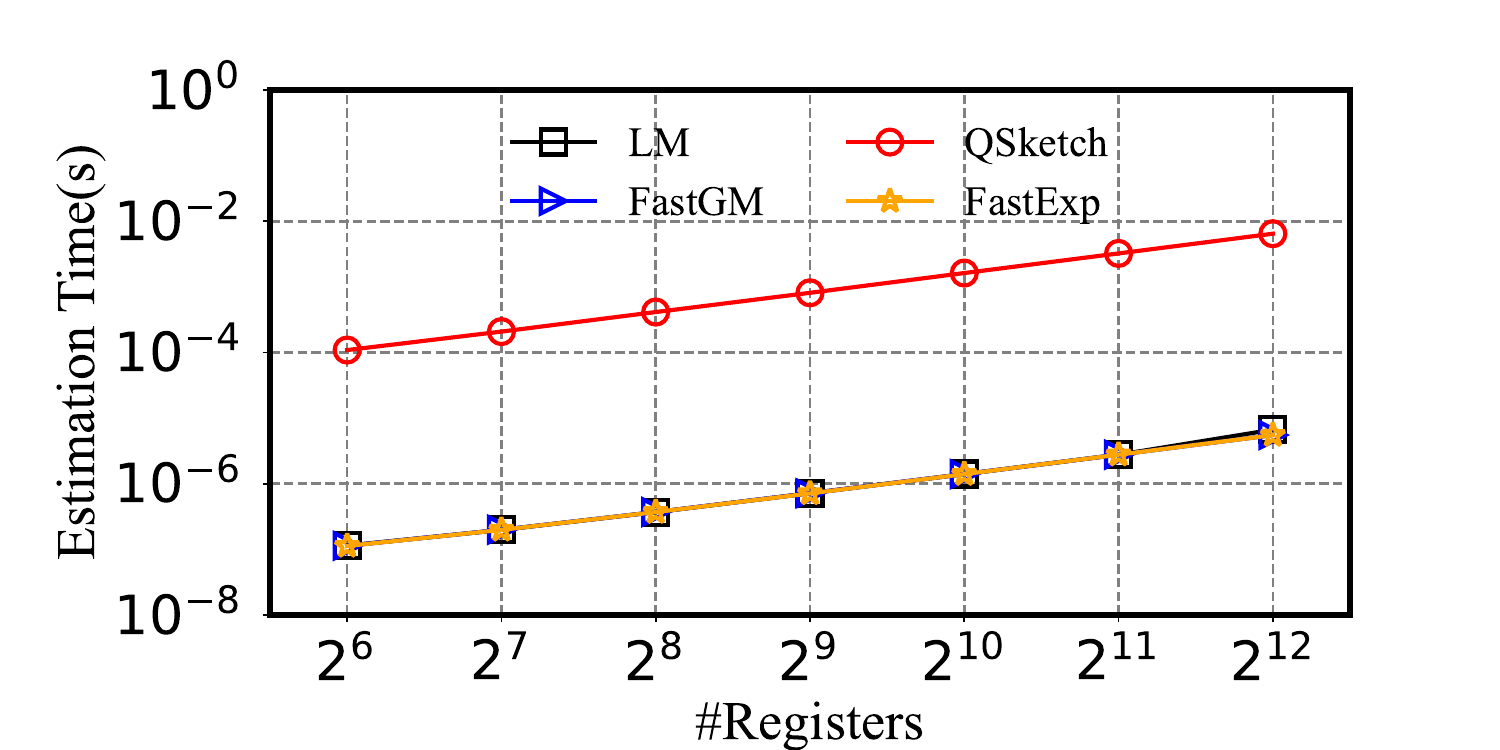}}
  	\subfigure[Uniform-10k]{\includegraphics[width=0.329\textwidth,trim=26 5 60 30,clip]{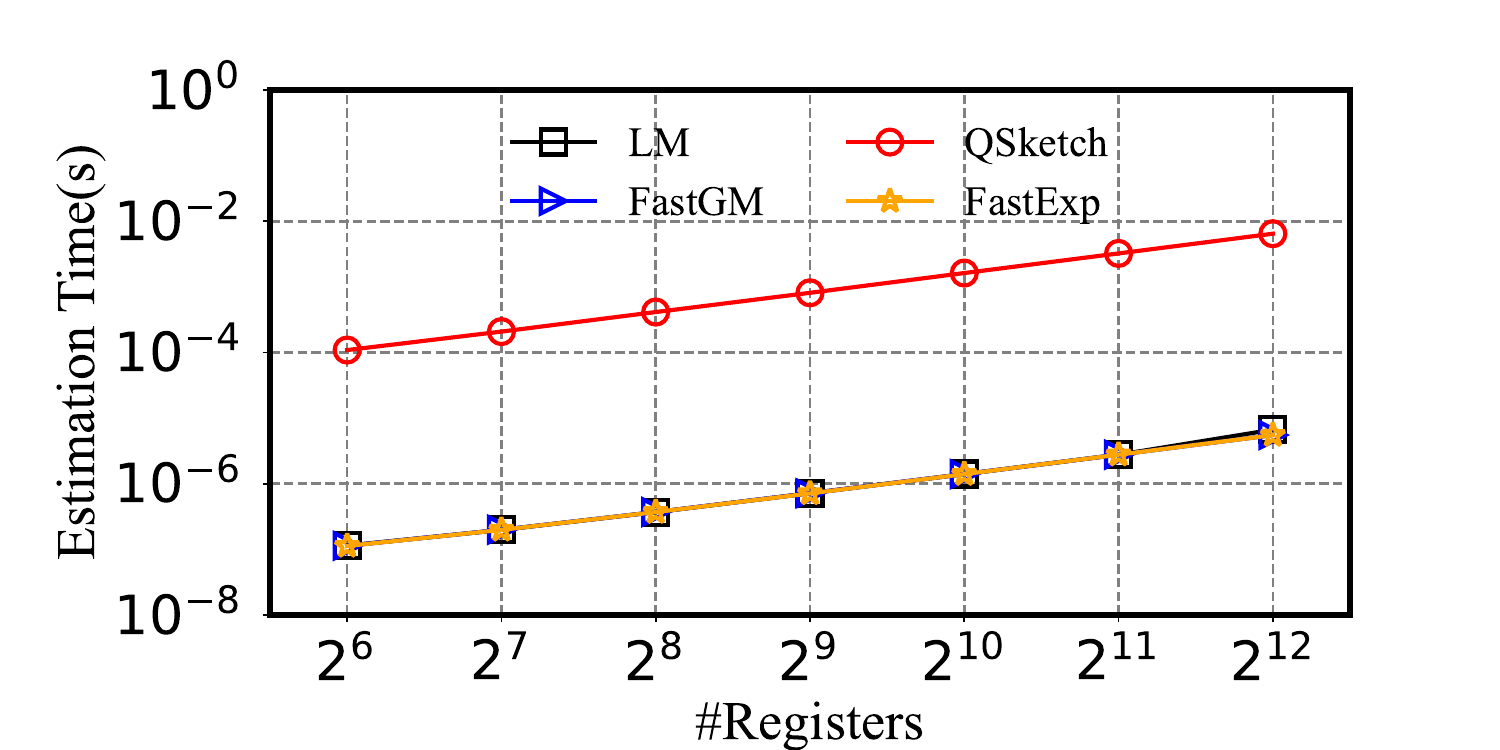}}
    \vspace{-5mm}
    \caption{Estimation time of all methods under different numbers of registers on synthetic datasets.}
    \vspace{-3mm}
    \label{figs:estimation_time}
\end{figure*}

Figure~\ref{figs:synthetic_update_time} shows the experimental results of update throughput on synthetic datasets with three different distributions.
Remarkably, the update throughput exhibits similar trends across all three distributions. 
Overall, QSketch-Dyn emerges as the superior performer among all competitors, a trend consistent with the results observed on real-world datasets. 
Specifically, the update throughput for QSketch-Dyn is approximately 10 and 100 times shorter compared to FastGM and LM, respectively. 
Figure~\ref{figs:estimation_time} shows the estimation time of all methods on three synthetic datasets.
We omit similar results on other datasets.
The estimation time of LM, FastGM, and FastExp Sketch is only related to the number of registers in the sketch.
QSketch needs several iterations for convergence, which costs more time.
Fortunately, in practical application scenarios, 
the estimation procedure may happen much less frequently than the update procedure,
and the absolute estimation time of QSketch is only 0.01s when using $4,096$ registers, which is acceptable.
Besides, QSketch-Dyn keeps track of the weighted cardinality on the fly,
and it does not need an estimation procedure.

\section{Related Work} \label{sec:related}

\subsection{Cardinality Estimation}
Harmouch et al.~\cite{harmouch2017cardinality} give a comprehensive review of existing sketch methods of estimating the cardinality. 
Whang et al.~\cite{Whang1990} introduce the LPC sketch using random hash functions for element mapping.
Various enhancements to LPC's estimation range were later proposed \cite{Estan2003, ChenJASA2011}.
Flajolet and Martin~\cite{Flajolet1985} develop the FM sketch, which was subsequently refined through methods like LogLog~\cite{Durand2003}, HyperLogLog~\cite{FlajoletAOFA07}, RoughEstimator~\cite{KanePODS2010}, and HLL-TailCut+~\cite{XiaoZC17}, reducing register size and employing multiple registers.
Giroire et al.~\cite{GiroireDAMNew2009} develop a sketching method \emph{MinCount} (also known as bottom-$k$ sketch~\cite{CohenPODC2007}) which stores the $k$ minimum hash values of elements in the set.
Ting~\cite{TingKDD2014} develops a martingale-based estimator to improve the accuracy of the above sketch methods such as LPC, HyperLogLog, and MinCount.
Chen et al.~\cite{Chen2013} extend HyperLogLog to sliding windows.
Besides sketch methods, two sampling methods~\cite{FlajoletComputing1990,GibbonsPVLDB2001} are also proposed for cardinality estimation.
Recently, considerable attention~\cite{Zhao2005, Yoon2009, WangTIFS2012, XiaoSIGMETRICS2015} has been given to developing fast sketch methods for monitoring the cardinalities of network hosts over high-speed links.
Ting~\cite{TingKDD2016} developed methods to estimate the cardinality of set unions and intersections from MinCount sketches.
Cohen et al.~\cite{CohenKDD2017} developed a method combining MinHash and HyperLogLog to estimate set intersection cardinalities.
Karppa et al. \cite{karppa2022hyperlogloglog} optimized HyperLogLog by decomposing register values into a base value and an offset vector, leading to more efficient storage.
Very recently, Wang et al.~\cite{wang2024half} proposed a method named Half-Xor to handle element deletions in cardinality estimation.

\subsection{Weighted Cardinality Estimation}
To estimate the weighted cardinality, 
Considine et al. \cite{considine2004approximate} used binary representations to represent integer weights, 
which is not efficient for elements with large weights.
Cohen et al. \cite{cohen2008tighter} proposed a weighted estimator based on bottom-$k$ sketches.
However, bottom-$k$ sketches require maintaining a sorted list of the $k$ smallest values, which needs more updating time and memory usage.
Recently, Lemiesz \cite{lemiesz2021algebra} presented a method that maps each element to $m$ exponential distributed variables.
Thus it needs $O(m)$ time to process an incoming element, which is infeasible for high-speed streams.
Therefore, Zhang et al. \cite{zhang2023fast} proposed FastGM to accelerate \cite{lemiesz2021algebra}.
FastGM generates these exponential variables in ascending order and stops the generation in advance
if the generated value is greater than the maximal value in current registers.
As a recent simultaneous work, FastExpSketch~\cite{lemiesz2023efficient} shares the same idea with FastGM.
However, these methods store generated values with 32-bit or 64-bit floating-point registers.
When we need a large $m$ to achieve better accuracy or there are many different streams, 
it is memory-intensive for devices with limited computational and storage resources.
In addition, these methods need $O(m)$ to estimate weighted cardinality,
and it is not efficient for them to provide anytime-available estimation for real-time applications.

\section{Conclusions and Future Work} \label{sec:conclusions}
This paper introduces QSketch, a memory-efficient sketching technique that leverages quantization methods to transform continuous register values into discrete integers. 
Unlike traditional sketching approaches which allocate 64 bits per register, our QSketch achieves comparable performance using only 8 bits per register. 
The QSketch experiences a worst-case time of $O(m\cdot n)$ where the weights of elements increase over time,
and it needs to solve an MLE problem through the Newton-Raphson method, which introduces extra computational overhead.
Therefore, we further capitalize on the dynamic nature of sketches by proposing QSketch-Dyn, which enables real-time monitoring of weighted cardinality. 
This enhanced method reduces estimation errors and maintains a constant time complexity for updates. 
We validate our approach through experiments conducted on both synthetic and real-world datasets. 
The results demonstrate that our novel sketching approach outperforms existing methods by approximately $30\%$ while consuming only one-eighth of the memory. 
In the future, we aim to explore weighted cardinality in streaming scenarios, particularly focusing on handling element deletions and elements with negative weights.

\section*{Acknowledgment}
The authors thank the reviewers for their comments and suggestions. 
This work was supported by the National Natural Science Foundation of China (U22B2019, 62372362, 62272372, 62272379).

\newpage
\balance
\bibliographystyle{ACM-Reference-Format}
\bibliography{ref1214}

\newpage
\appendix
\section{Appendxix}\label{sec:appendix}

\subsection{Fisher-Yates shuffle}\label{alg:fys}

The Fisher-Yates shuffle is commonly utilized to generate an unbiased permutation step-by-step.
The pseudocode of the shuffle is summarized in Algorithm~\ref{alg:fisher}.
Note that after the $i$-th step of the Fisher-Yates shuffle,
the first $i$ positions of the permutation are computed and fixed.
\begin{algorithm}[h]
\caption{Fisher-Yates shuffle.}
	\SetKwFunction{Swap}{Swap}
	\SetKwFunction{RandInt}{RandInt}
	\SetKwInOut{Input}{input}
	\SetKwInOut{Output}{output}
	\Input{An initial permutation $[\pi_1, \ldots, \pi_m]$.}
	\Output{Shuffled permutation $[\pi'_1, \ldots, \pi'_m]$.}
	\BlankLine
	$[\pi_1, \ldots, \pi_m] \gets [1, \ldots, m]$ \;
	\ForEach {$i \in \{1,\ldots, m\}$}{
            $k \gets$ \RandInt$(i, m)$ \;
    	\Swap$(\pi_k, \pi_i)$\;
	}
\end{algorithm}\label{alg:fisher}

\subsection{Proof of Theorem~\ref{theorem1}}\label{appendix:th1}
According to Equation~(\ref{eq:pri}), 
the probability that a register value $R[j]\le r_\text{min}$ or $R[j]\ge r_\text{max}$ is computed as,
\[
\begin{aligned}
    Pr(R[j]\le r_\text{min}) &= \sum_{r=-\infty}^{r_\text{min}} Pr(R[j]=r) \\
                             &= \int_{2^{-(r_\text{min}+1)}}^{+\infty} C_\Pi\cdot e^{-C_{\Pi} x} dx \\
                             &= e^{-C_\Pi\cdot 2^{-(r_\text{min}+1)}},
\end{aligned}
\]
and
\[
\begin{aligned}
    Pr(R[j]\ge r_\text{max}) &= \sum_{r_\text{min}}^{r=+\infty} Pr(R[j]=r) \\
                             &= \int_{0}^{2^{-r_\text{max}}} C_\Pi\cdot e^{-C_{\Pi} x} dx \\
                             &= 1 - e^{-c_\Pi\cdot 2^{-r_\text{max}}}.
\end{aligned}
\]
By setting $Pr(R[j]\le r_\text{min}) < \epsilon$ and $Pr(R[j]\ge r_\text{max}) < \epsilon$ simultaneously for above formulas,
we have
\[
- 2^{(r_\text{min}+1)}\cdot\ln{\epsilon} < C_\Pi < - 2^{r_\text{max}}\cdot\ln{(1-\epsilon)},
\]
which is the conclusion of the theorem.

\subsection{Proof of Theorem~\ref{theorem4}}\label{appendix:th4}

\begin{figure}[h]
    \centering
    \includegraphics[width=0.49\textwidth]{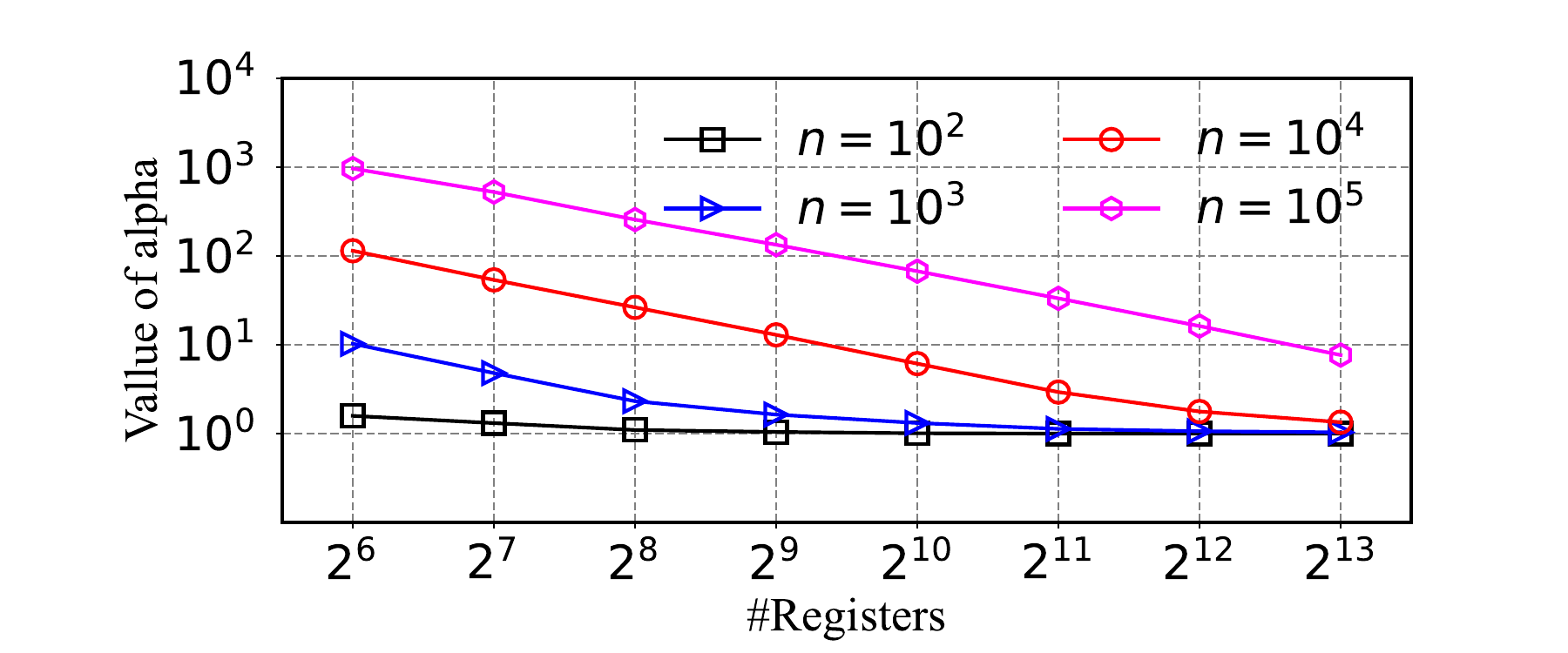}
    \caption{The value of alpha with different $n$ and $m$ on uniform distribution.}
    \label{fig:alpha}
\end{figure}

\begin{figure*}[!t]
    \centering
 	\subfigure[Accuracy]{\includegraphics[width=0.4\textwidth]{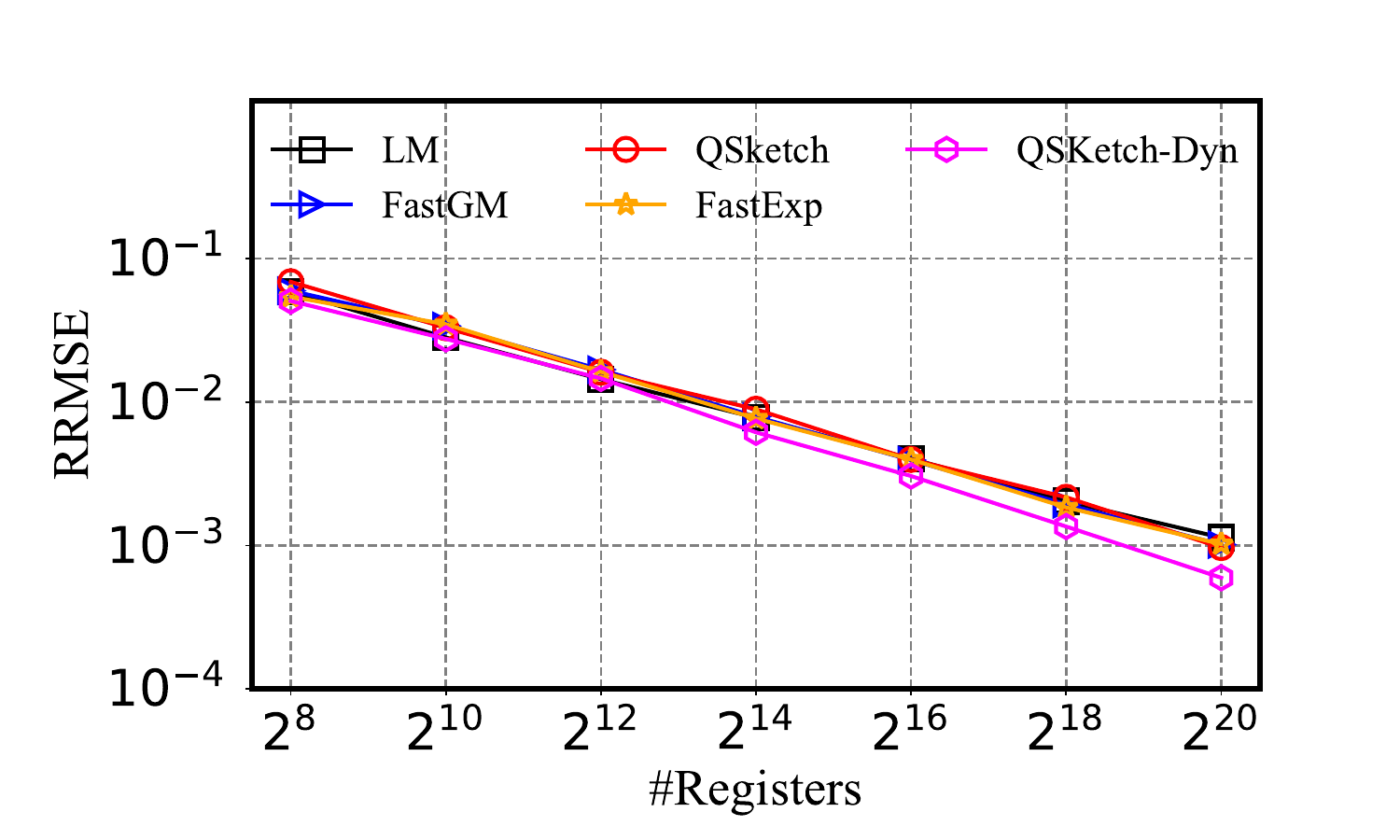}}
 	\subfigure[Update Throughput]{\includegraphics[width=0.4\textwidth]{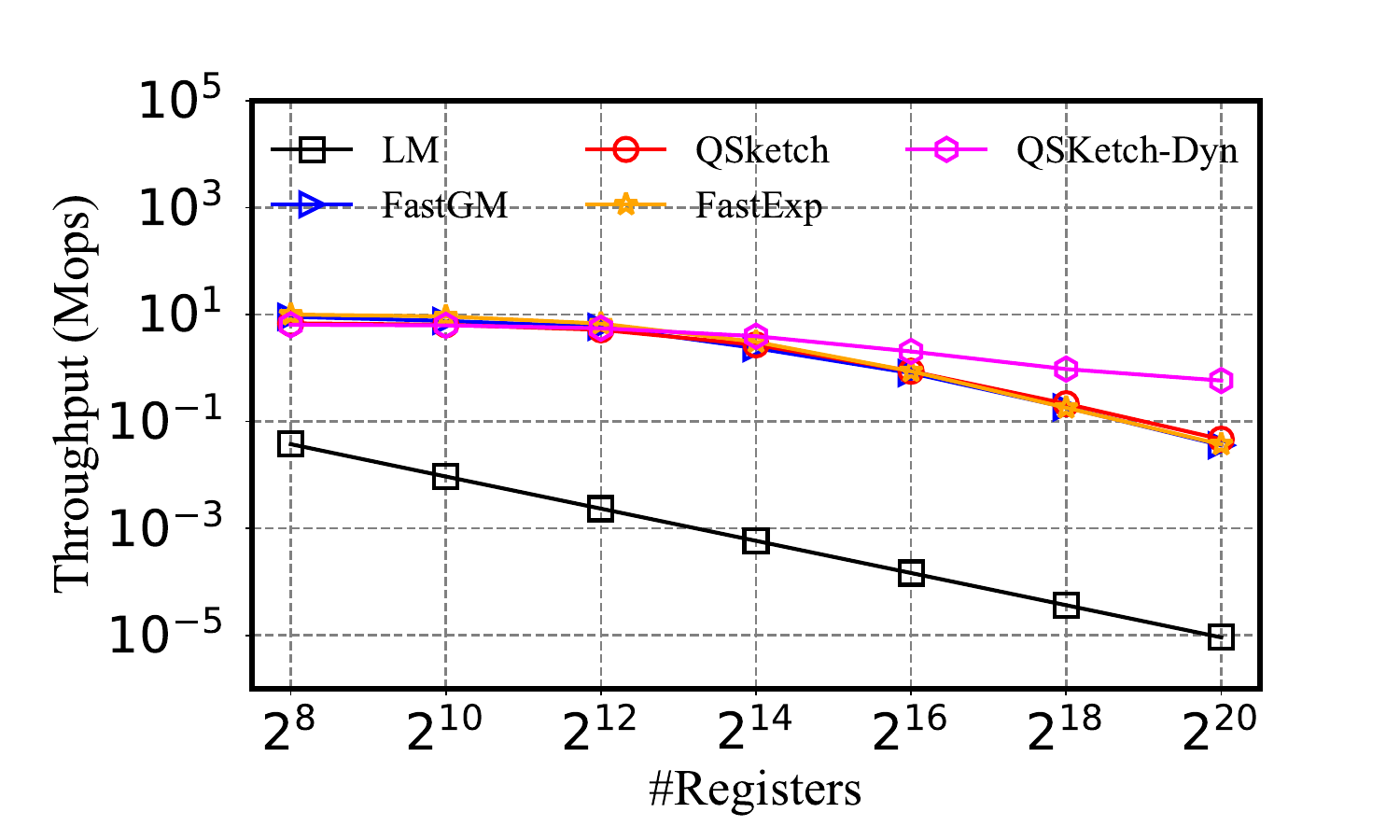}}
    \vspace{-5mm}
    \caption{Experiments on CAIDA dataset under a different number of registers.}
    \label{fig:caida}
    \vspace{-4mm}
\end{figure*}

Given the data stream $\Pi$ the set of timestamps that each element appears in the stream for the first time $T_s^{(t)}$, we first derive the expectation.
Let $\textbf{1}(R^{(t)} \neq R^{(t-1)})$ demote an indicator of whether the state of sketch $R$ has changed at time $t$, i.e., $\textbf{1}(R^{(t)} \neq R^{(t-1)}) = 1$ for $R^{(t)} \neq R^{(t-1)}$ and $0$ otherwise.
We note that $q_R^{(t)}$ only depends on $R^{(t-1)}$ and then we have 
\begin{equation*}
    \mathbb{E}[\textbf{1}(R^{(t)} \neq R^{(t-1)}) | q_R^{(t)}] = \mathbb{E}[\textbf{1}(R^{(t)} \neq R^{(t-1)}) | R^{(t-1)}] = q_R^{(t)}.
\end{equation*} 
From the linearity of expectation and the law of total expectation, we have
\begin{equation*}
\begin{aligned}
    &\mathbb{E}[\hat{C}^{(t)}_{\Pi}] = \mathbb{E}[\mathbb{E}[\hat{C}^{(t)}_{\Pi} | R^{(t-1)}]] \\
    &=\mathbb{E}[\mathbb{E}[\hat{C}^{(t-1)}_{\Pi} | R^{(t-1)}] + \mathbb{E}[\frac{\textbf{1}(R^{(t)} \neq R^{(t-1)})}{q_R^{(t)}} w^{(t)} | R^{(t-1)}]] \\
    &=\mathbb{E}[\hat{C}^{(t-1)}_{\Pi}] + w^{(t)} =  C^{(t)}_{\Pi}.
\end{aligned}
\end{equation*}
For the variance, we have 
\begin{equation*}
\begin{aligned}
\text{Var}[\hat{C}^{(t)}_{\Pi}] = \mathbb{E}[(\hat{C}^{(t)}_{\Pi})^2] - \mathbb{E}[\hat{C}^{(t)}_{\Pi}]^2 = \mathbb{E}[(\hat{C}^{(t)}_{\Pi})^2] - (C^{(t)}_{\Pi})^2.
\end{aligned}
\end{equation*}
Following a similar derivation with the expectation, we have
\begin{equation*}
\begin{aligned}
\mathbb{E}[(\hat{C}^{(t)}_{\Pi})^2] &= \mathbb{E}[\mathbb{E}[(\hat{C}^{(t)}_{\Pi})^2 | R^{(t-1)}]], 
\end{aligned}
\end{equation*}
where $\mathbb{E}[(\hat{C}^{(t)}_{\Pi})^2 | R^{(t-1)}] $ is computed as
\begin{equation*}
\begin{aligned}
&\mathbb{E}[(\hat{C}^{(t)}_{\Pi})^2 | R^{(t-1)}] \\
=& (\hat{C}^{(t-1)}_{\Pi})^2 + \frac{2 w^{(t)} \hat{C}^{(t-1)}_{\Pi}}{q_R^{(t)}} \mathbb{E}[\textbf{1}(R^{(t)} \neq R^{(t-1)}) | R^{(t-1)}] +\\
&( \frac{w^{(t)}}{q_R^{(t)}} )^2 \mathbb{E}[(\textbf{1}(R^{(t)} \neq R^{(t-1)}))^2 | R^{(t-1)}] \\
=&(\hat{C}^{(t-1)}_{\Pi})^2 + 2 w^{(t)} \hat{C}^{(t-1)}_{\Pi} + \frac{(w^{(t)})^2}{q_R^{(t)}} . 
\end{aligned}
\end{equation*}
Then, we have
\begin{equation*}
\begin{aligned}
\mathbb{E}[(\hat{C}^{(t)}_{\Pi})^2] &= \mathbb{E}[(\hat{C}^{(t-1)}_{\Pi})^2 + 2 w^{(t)} \hat{C}^{(t-1)}_{\Pi} + \frac{(w^{(t)})^2}{q_R^{(t)}}] \\
=&\mathbb{E}[(\hat{C}^{(t-1)}_{\Pi})^2] + 2 w^{(t)} \mathbb{E}[\hat{C}^{(t-1)}_{\Pi}] + \mathbb{E}[\frac{(w^{(t)})^2}{q_R^{(t)}}] \\
=&\sum\limits_{i \in T_s^{(t)}} \mathbb{E}[\frac{(w^{(i)})^2}{q_R^{(i)}}] + 2 \sum\limits_{i,j \in T_s^{(t)} \wedge i \neq j} w^{(i)}w^{(j)}.
\end{aligned}
\end{equation*}
Finally, we obtain
\begin{equation*}
\begin{aligned}
\text{Var}[\hat{C}^{(t)}_{\Pi}] = \mathbb{E}[(\hat{C}^{(t)}_{\Pi})^2] - \mathbb{E}[\hat{C}^{(t)}_{\Pi}]^2 = \sum\limits_{i \in T_s^{(t)}} (w^{(i)})^2 \mathbb{E}[\frac{1 - q_R^{(i)}}{q_R^{(i)}}].
\end{aligned}
\end{equation*}
An exact expression for $\mathbb{E}[\frac{1 - q_R^{(i)}}{q_R^{(i)}}]$ can be derived with the probability $P(R^{(t)}[1]=r_1, \ldots, R^{(t)}[m]=r_m | \Pi)$. However, it is too complex to analyze. 
Hence, we try to approximate the variance via the empirical analysis.
\begin{equation*}
\begin{aligned}
\text{Var}[\hat{C}^{(t)}_{\Pi}] <& w^{(t)}_{\max} \sum\limits_{i \in T_s^{(t)}} \mathbb{E}[\frac{w^{(i)}}{q_R^{(i)}}] - w^{(t)}_{\max} C^{(t)}_{\Pi}\\
=&(\alpha_{dis}(n, m) - 1) w^{(t)}_{\max} C^{(t)}_{\Pi},
\end{aligned}
\end{equation*}
where $w^{(t)}_{\max} = \max_{i \in T_s^{(t)}} w^{(i)}$ and $\alpha_{dis}(n, m)$ is a function of number of elements $n$ and number of registers $m$.

As an illustration, we consider a uniform distribution from the interval $(0, 1)$. Figure~\ref{fig:alpha} depicts the variation in the function's value across different values of $m$ and $n$.
As a result, we can get an upper bound of the variance together with the weighted cardinality estimation.

\subsection{Results on Large-Scale Dataset CAIDA}

We further conduct experiments on the CAIDA~\cite{caida} dataset,
which consists of streams of anonymized IP items collected from high-speed monitors by CAIDA in 2018.
A 1-minute CAIDA network traffic trace contains about 27M packets.
For each packet, we consider the tuple (source IP, target IP) as the identifier of the element $\mathrm{e}$,
and the packet size as the weight of the element $\mathrm{e}$.
Experimental results are summarized in Figure~\ref{fig:caida}.
We vary the number of registers in each sketch $m\in\{2^{8}, 2^{10}, 2^{12}, 2^{14}, 2^{16}, 2^{18}, 2^{20}\}$,
and evaluate the RRMSE as well as the update throughput.
All methods achieve better estimation accuracy when using more registers at the cost of lower update throughput.
Besides, we observe that:

\noindent $\bullet$ \textbf{QSketch achieves similar estimation accuracy as LM, FastGM, and FastGM,
and QSketch-Dyn performs best among all methods.}
This is consistent with previous results.
For example, when using $m=2^{20}$ registers, the estimation accuracy of QSketch-Dyn is about twice that of other methods.

\noindent $\bullet$ \textbf{The update throughput for QSketch-Dyn remains nearly consistent across varying numbers of
registers and is higher than other methods.}
This is also consistent with previous results.
For example, 
when use $m=2^{19}$ registers, 
the update throughput of QSketch-Dyn is about 1 Mops,
while the update throughput of FastGM, QSketch, and FastExp Sketch is only about 0.2 Mops,
which means that QSketch-Dyn is about $5 \times$ faster.
\end{document}